\documentclass[twocolumn,pre,english]{revtex4-1}
\usepackage[T1]{fontenc}
\usepackage{babel}
\usepackage{array}
\usepackage{float}
\usepackage{booktabs}
\usepackage{textcomp}
\usepackage{amsmath}
\usepackage{amsthm}
\usepackage{amssymb}
\usepackage{graphicx}
\usepackage{esint}
\usepackage{xcolor}
\usepackage{color}
\usepackage[unicode=true,pdfusetitle,
 bookmarks=true,bookmarksnumbered=false,bookmarksopen=false,
 breaklinks=false,pdfborder={0 0 1},backref=false,colorlinks=false]
 {hyperref}
 
\usepackage[sort&compress]{natbib}

\usepackage{silence}
\makeatletter

\providecommand{\tabularnewline}{\\}

\makeatother

\begin{document}
%\linenumbers

\title{Beyond Nearest-neighbour Universality of Spectral Fluctuations in Quantum Chaotic and Complex Many-body Systems}
\author{Debojyoti Kundu} \email{debojyoti.kundu.physics@gmail.com [Present Address : International Centre for Theoretical Sciences (ICTS) - TIFR, Bengaluru, India.]}
\author{Santosh Kumar} \email{Corresponding Author : skumar.physics@gmail.com}
\author{Subhra Sen Gupta} \email{Corresponding Author : subhro.sengupta@gmail.com}
\affiliation{\textrm{\textit{Department of Physics, Shiv Nadar Institution of Eminence (SNIoE), Gautam Buddha Nagar, Uttar Pradesh 201314, India}}} 
\begin{abstract}
Discerning chaos in quantum systems is an important problem as the usual route of Lyapunov exponents in classical systems is not straightforward in quantum systems. A standard route is the comparison of statistics derived from model physical systems to those from random matrix theory (RMT) ensembles, of which the most popular is the nearest-neighbour-spacings distribution (NNSD), which almost always shows good agreement with chaotic quantum systems. However, even in these cases, the long-range statistics (like number variance, spectral rigidity etc.), which are also more difficult to calculate, often show disagreements with RMT. As such, a more stringent test for chaos in quantum systems, via an analysis of intermediate-range statistics is needed, which will additionally assess the extent of agreement with RMT universality. In this paper, we deduce the effective level-repulsion parameters and the corresponding Wigner-surmise-like results of the next-nearest-neighbor spacing distribution (nNNSD) for integrable systems (semi-Poissonian statistics) as well as the three classical quantum-chaotic Wigner-Dyson regimes, by stringent comparisons to numerical RMT models and benchmarking against our exact analytical results for $3\times 3$ Gaussian matrix models, along with a semi-analytical form for the nNNSD in the Orthogonal-to-Unitary symmetry crossover. To illustrate the robustness of these RMT based results, we test these predictions against the nNNSD obtained from quantum chaotic models as well as disordered lattice spin models. This reinforces the Bohigas-Giannoni-Schmit and the Berry-Tabor conjectures, extending the associated universality to longer range statistics. In passing, we also highlight the equivalence of nNNSD in the apparently distinct Orthogonal-to-Unitary and diluted-Symplectic-to-Unitary crossovers.
\end{abstract}

\maketitle

\section{Introduction}

Understanding and quantifying chaos in the quantum domain is a complex and topical problem~\cite{Haake-Chaos_book,Reichl_chaos_book,Stockmann_chaos_book}. The standard approach of testing for hyper-sensitivity to initial conditions via positive Lyapunov exponents in the classical domain~\cite{Ott_Chaos_Book,Alligood-Sauer-Yorke_Chaos_Book}, does not work for quantum systems in any straightforward way due to ambiguity in the very concept of {\em trajectory}, in view of the uncertainty principle. The distinction of integrable systems vis-a-vis chaotic systems in quantum mechanics is a tricky problem~\cite{Quantum-Integrability_Papers}, but a standard prescription has emerged over the years due to the seminal works of Berry and Tabor (Berry-Tabor conjecture)~\cite{Berry-Tabor_conjecture_paper_1977.0140} on one hand, and Bohigas, Giannoni and Schmit (BGS conjecture)~\cite{BGS-conjecture} on the other. These conjectures relate the local spectral fluctuations (e.g. NNSD) for quantum integrable systems to Poissonian statistics, and those of quantum chaotic systems to the statistics of the three standard Random Matrix Theory (RMT) Wigner-Dyson classes, further sub-classified according to the presence/absence of various anti-unitary (time-reversal) and unitary (spatial) symmetries~\cite{Dyson-3fold,Dyson-RMT}, respectively.

RMT was first introduced in Physics by Wigner to understand the statistical properties of neutron scattering spectra of heavy nuclei~\cite{Wigner-before-RMT-1951,Wigner-Conference,Wigner-book-1957} and the mathematical framework and its connections to physical systems and their symmetries were further elucidated by the work of Wigner, Dyson, Porter, Mehta, Pandey and others~\cite{Wigner-RMT-2,Wigner-RMT-3,Wigner-RMT-4,Wigner-Siam-review,Mehta-1960,Dyson-RMT,DysonBM,Porter_nNNSD_GOE_1963,Porter_nNNSD_GUE_1963,Porter_nNNSD_book_1965,MehtaPandey1983,PandeyMehta1983}. Over the years, RMT has seen an unprecedented number and variety of applications in Physics as well as in many inter-disciplinary subjects, ranging from disordered quantum many-body systems~\cite{Mehta-1960,Mehta-Book,Haake-Chaos_book,Reichl_chaos_book,Stockmann_chaos_book,Wigner-Siam-review,Mehta_Dyson_paper,Mehta_Dyson_paper_2,T.GuhrReview,Beenakker_RMT_in_Quantum_Transport,Alhassid_quantum_dots}, to large complex atoms, molecules, and quantum chaotic systems \cite{Mehta-Book,Haake-Chaos_book,Beenakker_RMT_in_Quantum_Transport,Alhassid_quantum_dots}, as 	also in finance, atmospheric
science, medical science, and complex network theory \cite{RMT-application-1,RMT-application-2,RMT-application-3,RMT-application-7,RMT-application-4,RMT-application-5,RMT-application-6,RMT-Handbook,Dettmann_nNNSD_2017}.

In RMT, the study of the statistical behavior of the eigenspectra
and eigenstates are of primary concern. The study of the eigenspectra
involves the investigation of the \textit{short-range} and the \textit{long-range}
spectral correlation properties. The investigation of \textit{local}
(short-range) fluctuations in the eigenspectrum often involves the
analysis of two widely used measures: the \textit{nearest-neighbor
spacing distribution} (NNSD), which characterizes the probability
distribution of spacings between consecutive eigenvalues, and the
\textit{ratio distribution} (RD), which examines the distribution
of ratios of the consecutive level spacings \cite{Wigner-surmise-for-high-order-Abul-Magd,Avishai-Richert-PRB,Collura_Bose_Hubbard_2,Chavda-Kota-PLA-1,Iyer_Oganesyan_quasi-periodic_system,Modak's-paper_2014,Debojyoti_paper_1_2022}.
The \textit{spectral rigidity} (SD) and \textit{number variance} (NV)
measures are commonly employed to study the \textit{global} (long-range)
fluctuations in the eigenspectrum \cite{Santhanam_long_range,Long_range_Pandey_Puri,long_range_Sarika_Jalan,Magd_Unfolding_long_range,Bertrand_spin_chain_ratio_1,Debojyoti_paper_2_2023}.
In the context of almost all spectral fluctuation measures, excluding those involving ratios of spacings, one usually needs to go through an {\em unfolding procedure} which scales the mean level spacing to unity, in order to compare spectral statistics from diverse physical systems to standard RMT classes.
However, the unfolding procedure of the eigenvalues is not unique
\cite{Mehta-Book,Haake-Chaos_book,Avishai-Richert-PRB,Debojyoti_paper_1_2022}
and the long-range measures are extremely sensitive to the unfolding
procedure \cite{non-unfolding_long_range,Magd_Unfolding_long_range,Debojyoti_paper_2_2023}.
Also, the presence of \textit{finite-size effects} significantly affect
these long-range measures for physical systems. In this context, the
higher-order level spacing and ratio distribution studies may offer
a more effective and numerically straightforward method to handle
the \textit{beyond-short-range} spectral correlations \cite{Higher-order_finite-element_distributions_Duras,Higher-order_energy_level_spacing_distributions_D_Engel_1998,Wigner-surmise-for-high-order-Abul-Magd,santhanam_higher_order_ratio_nNNSD,Chavda_Distribution_of_Higher_Order_Spacing_Ratios,Wen-Jia_rao_Higher_order_level_spacing_nNNSD,Bhosale_higher_order_ratio_nNNSD}. 

The well-known Wigner-surmise results for NNSD of the three standard
symmetry classes (GOE, GUE, and GSE), obtained from the joint probability
distribution function (JPDF) of $2\times2$ Gaussian random matrices,
serve as excellent approximations to the large-dimension exact results
\cite{Wigner-book-1957,Mehta-1960,Mehta-Book,Haake-Chaos_book,Wigner-Siam-review}.
Similar results for RD, based on $3\times3$ matrix models, can also
be found in Refs. \cite{AtasPRL2013_ratio_1,Atas_2013_ratio_2,TKS2018,Ayana's_PhysRevE}.
To derive the exact Wigner-surmise-like results for higher-order level
spacing and ratio distributions, one needs to deal with the higher
dimensional JPDF's and the process become quite cumbersome. There
are some early as well as recent studies \cite{Higher-order_finite-element_distributions_Duras,Higher-order_energy_level_spacing_distributions_D_Engel_1998,Wigner-surmise-for-high-order-Abul-Magd,Bogomolny_semi_poissonian,santhanam_higher_order_ratio_nNNSD,Chavda_Distribution_of_Higher_Order_Spacing_Ratios,Wen-Jia_rao_Higher_order_level_spacing_nNNSD,Bhosale_higher_order_ratio_nNNSD,Bhosale_higher_order_ratio_2}
which generalize these higher-order distributions by considering phenomenological
and numerical fitting arguments.

In RMT, the Dyson index $\beta$ is known to serve as the measure
of the degree of level-repulsion in the eigenspectra of a system.
An uncorrelated sequence of energy levels follow the Poissonian statistics
with no level-repulsion ($\beta=0$). The eigenspectra of the three
Wigner-Dyson Gaussian ensembles, GOE, GUE, and GSE, possess finite
degrees of level-repulsion and the corresponding $\beta=1,$ $2$,
and $4$, respectively. In these cases, the above values signify the
number of independent \textit{real} parameters in a typical \textit{off-diagonal}
element of the original matrix, that may be tweaked to control the
degree of level-repulsion amongst its eigenvalues. For GOE, GUE and
GSE, the off-diagonal elements being real numbers, complex numbers,
and quaternions, thus have $1$, $2$ and $4$ independently variable
parameters, respectively. The higher-order level-spacing and ratio
distributions, may also be expected to follow the generalized \textit{Wigner-surmise-like}
distributions \cite{Higher-order_energy_level_spacing_distributions_D_Engel_1998,Wigner-surmise-for-high-order-Abul-Magd,santhanam_higher_order_ratio_nNNSD,Chavda_Distribution_of_Higher_Order_Spacing_Ratios,Wen-Jia_rao_Higher_order_level_spacing_nNNSD,Bhosale_higher_order_ratio_nNNSD}.
Therefore, one can identify \textit{effective} Dyson indices, $\beta'$,
corresponding to the degree of the level spacings of an ensemble. 

The main objective of this paper is to establish a more stringent test for chaos in quantum systems, via an analysis of spectral statistics beyond NNSD, specifically the next-nearest-neighbour spacing distribution (nNNSD), for both the {\em integrable} as well as the {\em quantum chaotic} regimes. This is achieved by rigorous comparisons to a variety of physical systems, some with classically chaotic counterparts and some disordered quantum models with no classical counterpart, thus also assessing the extent of agreement with RMT universality. We deduce this based on the fitting of
generalized Wigner surmise-like ansatzes for the nNNSD, to the eigenvalue statistics
of large random matrices. To this end, we perform
Monte-Carlo simulations of relevant random matrix models to determine the associated effective Dyson indices
($\beta'$), and contrasted with earlier studies~\cite{Wigner-surmise-for-high-order-Abul-Magd,santhanam_higher_order_ratio_nNNSD,Wen-Jia_rao_Higher_order_level_spacing_nNNSD,Bhosale_higher_order_ratio_nNNSD}.
%Then by employing relevant \textit{normalization} and \textit{unit
%mean level-spacing} conditions, we obtain the Wigner-surmise-like
%expressions for nNNSD's of integrable systems and the three Wigner-Dyson
%ensembles. 
Furthermore, for the GOE and GUE, we present exact analytical
expressions of the nNNSD's using the JPDF of eigenvalues
of the corresponding $3\times3$ RMT matrix models. Also,
using the JPDF of the $3\times3$ GOE-to-GUE crossover model, we derive
a \textit{semi-analytical} expression for the corresponding nNNSD
crossover. The physical models used to examine the universality of
these results include two single-body (non-interacting)
quantum chaotic systems, namely the \textit{quantum kicked rotor}
(QKR) \cite{Izrailev_first_paper_QKR,Casati_QKR,Izrailev_QKR_2,Pandey_Jaiswal_QKR_2019,Pandey_Puri_QKR_2020}
and the \textit{quarter Sinai billiard} \cite{sinai_billiard_1963,Sinai_billiard_1970,Santosh-Rohit2020}, on one hand, and 
two many-body (interacting) disordered quantum systems, specifically two {\em disordered spin-chain models}~\cite{Debojyoti_paper_1_2022,Debojyoti_paper_2_2023}, on the other. {\em This corroborates the universality of the Berry-Tabor and BGS conjectures beyond nearest-neighbour spectral fluctuations for integrable as well as quantum chaotic systems with various possible symmetries.} 
%referred to as model-I
%and model-II. The model-I includes a variety of terms like the isotropic
%Heisenberg interaction, Zeeman coupling to a random, inhomogeneous
%magnetic field, and the scalar spin-chirality interaction \cite{Debojyoti_paper_1_2022}.
%Along with the Heisenberg and the Zeeman coupling terms, the model-II
%also includes a random Ising term and an anisotropic and antisymmetric,
%but deterministic Dzyaloshinskii-Moriya (DM) term \cite{Debojyoti_paper_2_2023}. 

Towards the end, we also highlight the equivalence of nNNSD in the seemingly distinct GOE-to-GUE and {\em diluted}-GSE-to-GUE crossovers, where {\em diluted} refers to the case where both partners of all Kramers doublets are retained in the GSE spectra~\cite{Debojyoti_paper_2_2023}. This is illustrated via an analysis of both RMT crossover matrix models as well as the two disordered, interacting lattice spin models. 
%The first transition is the standard GOE-to-GUE
%crossover in the nNNSD. In the second crossover, we begin with the
%nNNSD of the diluted GSE distribution (discussed in Ref. \cite{Debojyoti_paper_2_2023}),
%which retains both the Kramers degenerate (KD) partners from the eigenspectrum
%and effectively becomes the NNSD of the standard GSE. Now, a finite
%value of the crossover parameter, lifts the KD, and increasing it
%further breaks the time-reversal symmetry, allowing the system to
%transit to the standard nNNSD of the GUE.
Despite having very different
physical symmetry settings in two distinct spin models, it will be
argued that both of them can we viewed, in a certain sense, as a transition
between the NNSD of the standard GSE, and the nNNSD of the canonical
GUE, based on the mathematical form of the distribution functions
alone.

The rest of this paper is presented as follows. In Sec. \ref{sec:METHODOLOGY-and-CALCULATIONS},
we deduce the effective Dyson indices and thereby obtain the generalized Wigner-surmise expressions for the nNNSD's in the context of the integrable case as well as the
three Wigner-Dyson ensembles, via a fitting procedure to Monte-Carlo simulations involving large-dimensional matrices.
In Sec. \ref{sec:Derivation-of-exact-nNNSD-results} (supplemented by Appendix \ref{Appendix}),
we consider the JPDF's of the eigenvalues of $3\times 3$
random matrix models following GOE and GUE, and deduce the exact analytical
results for the nNNSD's. These exact analytical results show good
agreement with the generalized Wigner-surmise expressions obtained in Sec. \ref{sec:METHODOLOGY-and-CALCULATIONS}.
Similarly, for the GOE-to-GUE nNNSD crossover, we derive a semi-analytical
form and employ its numerical realizations for examining
the limiting and intermediate cases. In Sec. \ref{sec:Study-of-nNNSD-in-Quantum-Chaotic-Systems},
we investigate the QKR system in its integrable and strongly chaotic
regimes (before and after breaking of the time-reversal symmetry),
and study its nNNSD statistics. We also study the nNNSD's of the tight-binding
(TB) Hamiltonian in both an {\em integrable rectangular billiard} and a
{\em chaotic quarter Sinai billiard system} by using the computational
package Kwant \cite{Groth_Kwant_quantum_transport_2014}. Here, we also introduce the aforementioned disordered spin-chain models (referred to as {\em model-I}
and {\em model-II}), and explore the robustness of the previously deduced
nNNSD results (in Sections \ref{sec:METHODOLOGY-and-CALCULATIONS} and \ref{sec:Derivation-of-exact-nNNSD-results}) by comparing them with the corresponding spectral statistics of these spin systems. Lastly, in Sec. \ref{sec:Interesting-features-in-crossovers},
we discuss the equivalence of the two apparently distinct nNNSD crossovers as mentioned above. The key findings and conclusions are summarized in
Sec. \ref{sec:Conclusions}.

\section{METHODOLOGY and CALCULATIONS for RMT Level Spacing Distributions\label{sec:METHODOLOGY-and-CALCULATIONS}}

The NNSD is the most studied short-range level correlation statistics,
which represents the statistical behavior of the local fluctuations
of an eigenspectrum. From the ordered sequence of the unfolded eigenvalues,
$\mathop{\tilde{\varepsilon}_{1}\leqslant\cdots\leqslant\tilde{\varepsilon}_{j}\leqslant\tilde{\varepsilon}_{j+1}\cdots}$,
we can define the nearest-neighbor level spacing as $s_{j}=(\tilde{\varepsilon}_{j+1}-\tilde{\varepsilon}_{j})$
\cite{Debojyoti_paper_1_2022}. The NNSD of uncorrelated energy
levels follows the Poissonian statistics, as represented by 
\begin{equation}
\ensuremath{P_{\mathrm{Poi}}(s)=\exp(-s)}.\label{eq:Poissonian_NNSD}
\end{equation}
 On the other hand, the NNSD of correlated levels is given by the
Wigner-surmise formula \cite{Mehta-Book,Haake-Chaos_book} 
\begin{equation}
P_{\mathrm{WD}}(s)=a_{\beta}s^{\beta}\exp(-b_{\beta}s^{2}),\label{eq:general_NNSD_formula}
\end{equation}
 where $\beta$ is the \textit{Dyson index}, whose significance is
described in the Introduction above. For example, $\beta=1$, $2$,
and $4$ corresponds to the Gaussian orthogonal ensemble (GOE), Gaussian
unitary ensemble (GUE), and Gaussian symplectic ensemble (GSE), respectively.
The constants $a_{\beta}$ and $b_{\beta}$, in Eq. (\ref{eq:general_NNSD_formula}),
can be derived from the normalization and unit mean-level spacing
conditions, 
\begin{equation}
\begin{array}{cc}
\intop_{0}^{\infty}P_{\mathrm{WD}}(s)ds=1; & \intop_{0}^{\infty}sP_{\mathrm{WD}}(s)ds=1\end{array}.\label{eq:normalization_conditions-1}
\end{equation}
 The analytical NNSD formulas for the three Gaussian ensembles along
with that for the Poissonian statistics are compiled in Table \ref{tab:NNSD_formulas_table}. 

Along similar lines, one can also consider higher order level spacings
and study their corresponding distributions, assuming that they can
also be represented by Wigner-surmise-like formulae with some \textit{effective}
$\beta$'s (denoted by $\beta^{\prime}$). The probability distribution
of the level spacings of a correlated spectra can be represented by
the generalized Wigner-surmise formula \cite{Porter_nNNSD_book_1965,Wigner-surmise-for-high-order-Abul-Magd}, 

\begin{equation}
P_{\mathrm{WD}}^{m}(w_{m})=a_{\beta'}w_{m}^{\beta'}\exp(-b_{\beta'}w_{m}^{2}),\label{eq:general_RMT_level_spacing_formula}
\end{equation}
 where $m$ is the order of the level spacings and $\beta'$ represents
the corresponding \textit{effective} Dyson index. Here, $w_{m}$ is
the $m$-th order level spacing of the unfolded eigenvalues {[}$(w_{m})_{j}=(\tilde{\varepsilon}_{j+m}-\tilde{\varepsilon}_{j})/m${]}
\cite{Wigner-surmise-for-high-order-Abul-Magd}. The $m=1$ case represents
the nearest-neighbor spacings $s$ ($\equiv w_{1}$) and $m>1$ represents
higher order spacings. The constants, $a_{\beta'}$ and $b_{\beta'}$,
can be derived from the corresponding normalization and unit mean-level
spacing conditions, 
\begin{equation}
\begin{array}{cc}
\intop_{0}^{\infty}P_{\mathrm{WD}}^{m}(w_{m})dw_{m}=1; & \intop_{0}^{\infty}w_{m}P_{\mathrm{WD}}^{m}(w_{m})dw_{m}=1\end{array}.\label{eq:normalization_conditions-2}
\end{equation}

One can also generalize the probability distribution of the level
spacings of an eigenspectrum, consisting of uncorrelated levels to
higher order spacings, $i.e.$ \cite{Bogomolny_semi_poissonian},
\begin{equation}
P_{\mathrm{Poi}}^{m}(w_{m})=c_{\beta'}w_{m}^{\beta'}\exp(-d_{\beta'}w_{m}),\label{eq:Poissonian_generalized_level_spacing_formula}
\end{equation}
where $\beta'=0,m=1$ represents the standard NNSD of the Poissonian
statistics {[}$P_{\mathrm{Poi}}(s)${]}, $c_{\beta'}$ and $d_{\beta'}$
can also be derived from the corresponding conditions as in Eq. (\ref{eq:normalization_conditions-2}). 

\begin{table}
\caption{Probability distribution of nearest-neighbor spacings for unfolded eigenvalues \cite{Mehta-Book,Haake-Chaos_book}.\label{tab:NNSD_formulas_table}}
\begin{tabular}{cc}
\hline  
{\footnotesize{}Type of distribution} & {\footnotesize{}NNSD Probability Density}\tabularnewline
\hline 
{\footnotesize{} Poissonian} & $P_{\mathrm{Poi}}(s)=\exp(-s)$\tabularnewline
{\footnotesize{}GOE} & {\footnotesize{}$P_{\mathrm{GOE}}(s)=(\pi s/2)\exp(-\pi s^{2}/4)$}\tabularnewline
{\footnotesize{}GUE} & {\footnotesize{}$P_{\mathrm{GUE}}(s)=(32s^{2}/\pi^{2})\exp(-4s^{2}/\pi)$}\tabularnewline
{\footnotesize{}GSE} & {\footnotesize{}$P_{\mathrm{GSE}}(s)=(2^{18}s^{4}/3^{6}\pi^{3})\exp(-64s^{2}/9\pi)$}\tabularnewline
\hline 
\end{tabular}
\end{table}

It is known that the \textit{next} nearest-neighbor spacing distribution
of the Orthogonal (GOE) class has the exact same analytical form as
the nearest-neighbor spacing distribution of the Symplectic (\textit{standard}
GSE) class \cite{Mehta-Book,Mehta_Dyson_paper_2}, $viz.$, 
\begin{equation}
\ensuremath{\mathop{P_{\mathrm{GOE}}^{2}(w_{2})=(2^{18}/3^{6}\pi^{3})w_{2}^{4}\exp(-64w_{2}^{2}/9\pi)}},\label{eq:nNNSD_of_GOE}
\end{equation}
where $(w_{2})_{j}=(\tilde{\varepsilon}_{j+2}-\tilde{\varepsilon}_{j})/2$
\cite{Avishai-Richert-PRB,long_range_Sarika_Jalan} is the suitably
normalized next nearest-neighbor level spacing ($m=2$) of the unfolded
eigenvalues. For other Wigner-Dyson ensembles, the analytical forms
of the Wigner-surmise-like nNNSD's {[}$P_{\mathrm{WD}}^{2}(w_{2})${]}
are not widely known.

In this work, we address this issue based on a two-pronged approach.
In the \textit{first approach}, we assume a modified Poissonian or
a modified Wigner-Dyson form for the nNNSD distribution functions
as given by Eqs. (\ref{eq:Poissonian_generalized_level_spacing_formula})
and (\ref{eq:general_RMT_level_spacing_formula}), respectively, for
the integrable and the quantum-chaotic cases, respectively. We then
extract a form for the constants involved using the corresponding
normalization conditions {[}Eq. (\ref{eq:normalization_conditions-2}){]},
which are still expressed in terms of the unknown index $\beta^{\prime}$.
This index is determined by detailed fittings to numerical diagonalization
derived data for large matrices, as described below. In the \textit{second
approach} (discussed in the next section), we deduce exact formulas
for the nNNSD distribution functions for the GOE and GUE classes based
on an analytical treatment using the three-dimensional JPDF's for
the eigenvalues.

\begin{figure}[h]
\begin{centering}
\includegraphics[scale=0.33]{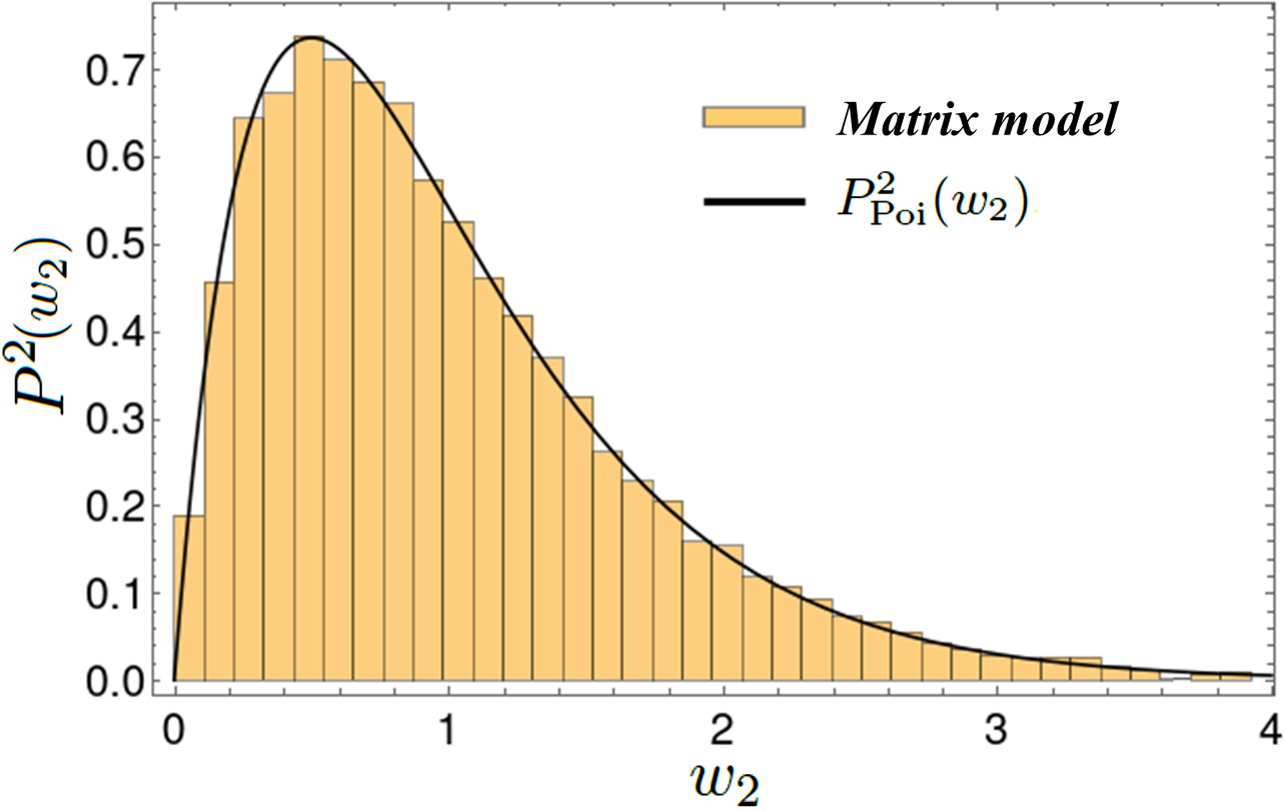}
\par\end{centering}
\caption{The nNNSD of the numerical matrix model matrix $\mathcal{H}_{\mathrm{Poi}}$
(dimension $n=15000$) which has uncorrelated eigenvalues. The spacing
distribution follows the semi-Poissonian distribution $P_{\mathrm{Poi}}^{2}(w_{2})$,
mentioned in the Table \ref{tab:nNNSD_formulas_table}. \label{fig:nNNSD_Poissonian}}
\end{figure}

\begin{figure}[h]
\begin{centering}
\includegraphics[scale=0.33]{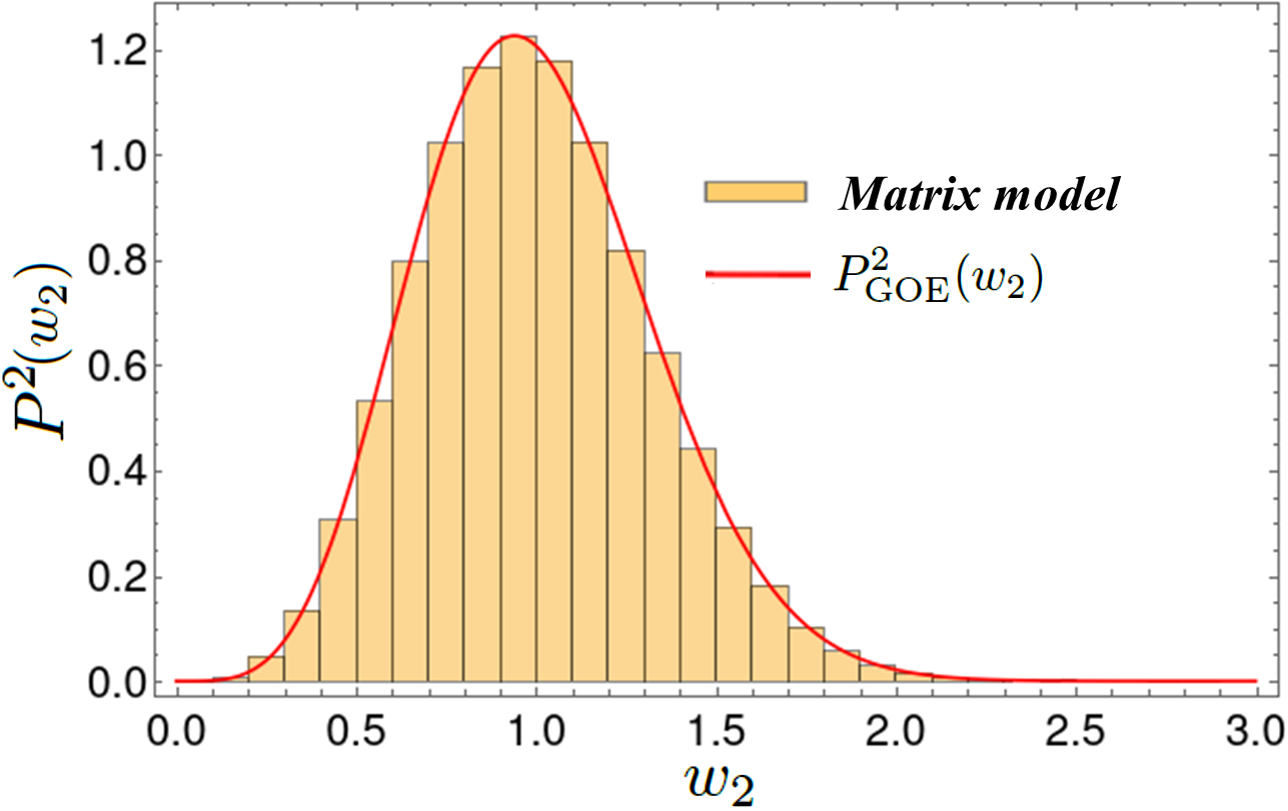}
\par\end{centering}
\caption{The nNNSD of the numerical random matrix model $\mathcal{H}_{\mathrm{GOE}}$
($n=10000$ and an ensemble of $\mathcal{M}=20$ matrices), which
perfectly follows the corresponding GOE distribution $P_{\mathrm{GOE}}^{2}(w_{2})$,
mentioned in the Table \ref{tab:nNNSD_formulas_table}. \label{fig:nNNSD_GOE}}
\end{figure}

\begin{table*}
\caption{Probability distribution of next-nearest-neighbor spacings for unfolded eigenvalues.\label{tab:nNNSD_formulas_table}}
\begin{tabular}{cc}
\hline  
{\footnotesize{}Type of distribution} & {\footnotesize{}nNNSD Probability Density}\tabularnewline
\hline 
{\footnotesize{} Semi-Poissonian} & $P_{\mathrm{Poi}}^{2}(w_{2})=4{w}_{2}\exp(-2{w}_{2})$\tabularnewline
{\footnotesize{}GOE} & {\footnotesize{}$P_{\mathrm{GOE}}^{2}(w_{2})=(2^{18}/3^{6}\pi^{3}){w}_{2}^{4}\exp(-64{w}_{2}^{2}/9\pi)$}\tabularnewline
{\footnotesize{}GUE} & {\footnotesize{}$P_{\mathrm{GUE}}^{2}(w_{2})=\left[5^{8}7^{8}\pi^{4}/(2^{40}\times 3)\right]{w}_{2}^{7}\exp(-5^{2}7^{2}\pi{w}_{2}^{2}/2^{10})$}\tabularnewline
{\footnotesize{}GSE} & {\footnotesize{}$P_{\mathrm{GSE}}^{2}(w_{2})=\left[3^{12}7^{14}11^{14}13^{14}\pi^{7}/(2^{157}\times 5)\right]{w}_{2}^{13}\exp(-3^{2}7^{2}11^{2}13^{2}\pi{w}_{2}^{2}/2^{22})$}\tabularnewline
\hline 
\end{tabular}
\end{table*}

To deduce the nNNSD of the integrable case (uncorrelated spectra),
we need to solve for the constants $c_{\beta'}$ and $d_{\beta'}$,
using Eq. (\ref{eq:Poissonian_generalized_level_spacing_formula})
{[}$P_{\mathrm{Poi}}^{m}(w_{m})${]} and the corresponding conditions
in Eq. (\ref{eq:normalization_conditions-2}). We get, \begin{align} \nonumber &c_{\beta'}=\frac{\Gamma(2+\beta')^{1+\beta'}}{\Gamma(1+\beta')^{2+\beta'}}=\frac{(1+\beta')^{(1+\beta')}}{\Gamma(1+\beta')}, \\ &d_{\beta'}=\frac{\Gamma(2+\beta')}{\Gamma(1+\beta')}=1+\beta', \label{eq:nNNSD_constants_for_Poissonian} \end{align}
where $\Gamma(z)$ is the \textit{Euler Gamma} function. Next, we
construct a numerical matrix ($\mathcal{H}_{\mathrm{Poi}}$) of dimension
$n=15000$ (large $n$ is considered to eliminate any possible finite-size
effects), whose eigenvalues are uncorrelated and NNSD follows Poissonian
statistics. To obtain the parameter $\beta'$, we perform the \textit{best
fit} of the probability distribution function $P_{\mathrm{Poi}}^{2}(w_{2})$
{[}considering $c_{\beta'}$ and $d_{\beta'}$ in the from of Eq.
(\ref{eq:nNNSD_constants_for_Poissonian}){]} to the nNNSD of the
unfolded spectra of the constructed matrix $\mathcal{H}_{\mathrm{Poi}}$.
We find $\beta'=1.006\approx1$, and using Eq. (\ref{eq:nNNSD_constants_for_Poissonian}),
the normalization constants become $c_{1}=4$ and $d_{1}=1$. This
functional form, $P_{\mathrm{Poi}}^{2}(w_{2})$, mentioned in the
Table \ref{tab:nNNSD_formulas_table}, is known as the \textit{semi-Poissonian
distribution.}\footnote{The NNSD of integrable systems follow Poissonian statistics, but the
higher-order spacing distributions (like nNNSD here) do not follow
Poissonian statistics, and in general, referred to as semi-Poissonian
distributions \cite{Bogomolny_semi_poissonian,Wen-Jia_rao_Higher_order_level_spacing_nNNSD}.
The nomenclature arises from the observation that, the NNSD of a new
sequence $\{\tilde{\varepsilon}'_{j}\}$, obtained by averaging neighboring
elements {[}$\tilde{\varepsilon}'_{j}=(\tilde{\varepsilon}_{j}+\tilde{\varepsilon}_{j+1})/2${]}
from an ordered Poissonian distributed sequence ($\mathop{\tilde{\varepsilon}_{1}\leqslant\cdots\leqslant\tilde{\varepsilon}_{j}\leqslant\tilde{\varepsilon}_{j+1}\leqslant\cdots}$),
follows the semi-Poissonian distribution $P_{\mathrm{Poi}}^{2}(w_{2})$.} In Fig. \ref{fig:nNNSD_Poissonian}, $P_{\mathrm{Poi}}^{2}(w_{2})$
is plotted along with the nNNSD \textit{histogram} of the corresponding
matrix model $\mathcal{H}_{\mathrm{Poi}}$.

\begin{figure}[h]
\begin{centering}
\includegraphics[scale=0.33]{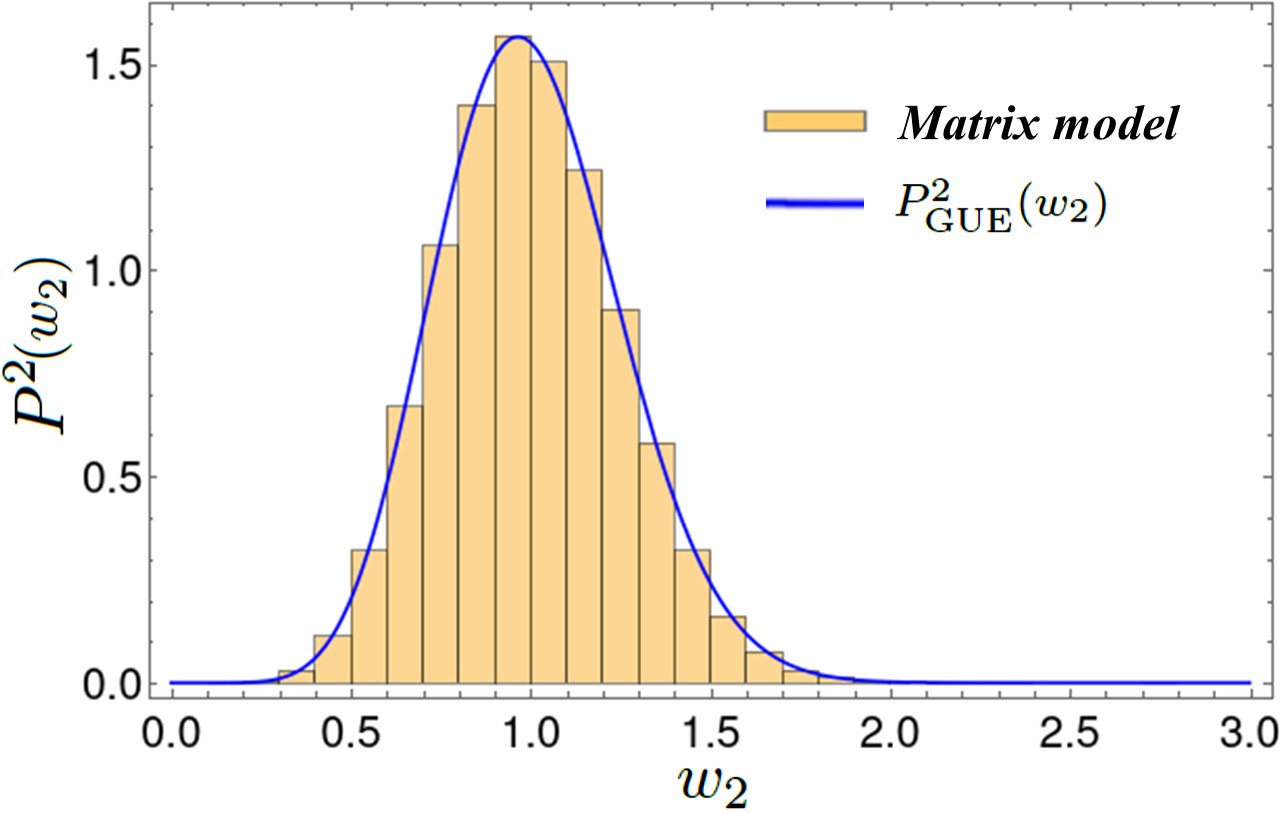}
\par\end{centering}
\caption{The nNNSD of the numerical random matrix model $\mathcal{H}_{\mathrm{GUE}}$
($n=40000$ and $\mathcal{M}=5$), which perfectly follows the deduced
GUE distribution $P_{\mathrm{GUE}}^{2}(w_{2})$, mentioned in the
Table  \ref{tab:nNNSD_formulas_table}. \label{fig:nNNSD_GUE}}
\end{figure}

\begin{figure}[H]
\begin{centering}
\includegraphics[scale=0.33]{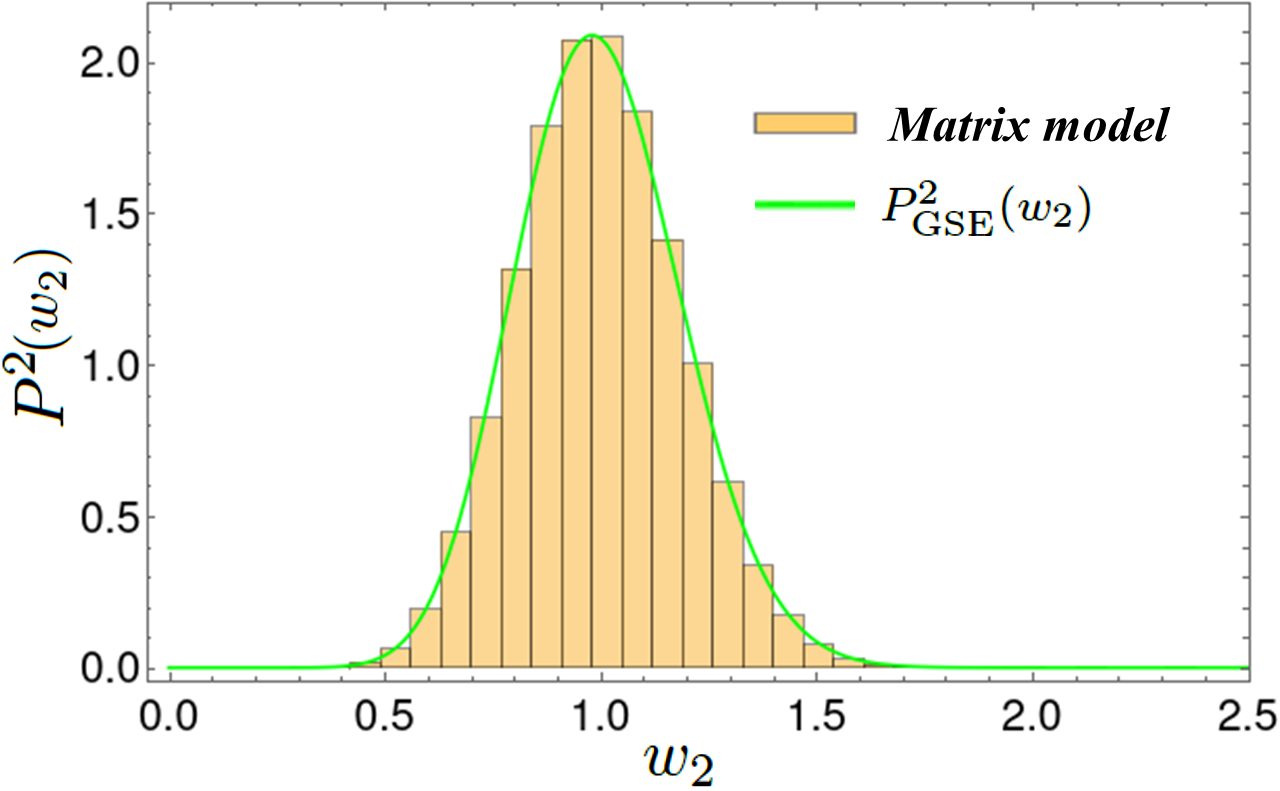}
\par\end{centering}
\caption{The nNNSD of the numerical random matrix model $\mathcal{H}_{\mathrm{GSE}}$
($n=40000$ and $\mathcal{M}=5$), which perfectly follows the deduced
GSE distribution $P_{\mathrm{GSE}}^{2}(w_{2})$, mentioned in the
Table \ref{tab:nNNSD_formulas_table}. \label{fig:nNNSD_GSE}}
\end{figure}

In the same manner, to find out the analytical expressions of the
nNNSD for the GOE, GUE and GSE classes, we solve for $a_{\beta'}$
and $b_{\beta'}$, using Eq. (\ref{eq:general_RMT_level_spacing_formula})
{[}$P_{\mathrm{WD}}^{m}(w_{m})${]} and the corresponding conditions
in Eq. (\ref{eq:normalization_conditions-2}). We get the generic
expressions: 
\begin{equation}
\begin{array}{cc}
a_{\beta'}=\frac{2(b_{\beta'})^{\left(\frac{1+\beta'}{2}\right)}}{\Gamma\left(\frac{1+\beta'}{2}\right)}, & b_{\beta'}=\frac{\left[\Gamma\left(1+\frac{\beta'}{2}\right)\right]^{2}}{\left[\Gamma\left(\frac{1+\beta'}{2}\right)\right]^{2}}\end{array}.\label{eq:nNNSD_constants_for_Gaussian_ensembles}
\end{equation}
 Then, we construct numerical Gaussian random matrices ($\mathcal{H}_{\mathrm{WD}}$)
of dimension $n$ ($=10000$, $40000$, and $40000$ for the GOE,
GUE, and GSE, respectively), from the respective RMT ensembles following
Wigner-Dyson symmetry classes. For these symmetry classes, we sample
the matrices using ensemble sizes of $\mathcal{M}=20,$ $5$, and
$5$, for the GOE, the GUE, and the GSE, respectively. The standard
unfolding procedure is carried out before computing the probability
distributions of the level spacings. For each case, we perform the
\textit{best fit} of the distribution $P_{\mathrm{WD}}^{2}(w_{2})$
{[}Eq. (\ref{eq:general_RMT_level_spacing_formula}) with $m=2$ and
considering $a_{\beta'}$ and $b_{\beta'}$ in the from of Eq. (\ref{eq:nNNSD_constants_for_Gaussian_ensembles}){]}
to the respective nNNSD of the eigenvalues of the RMT matrix models
and obtain the corresponding effective Dyson index, $\beta'$. As
expected, for the GOE class, using $\mathcal{H}_{\mathrm{GOE}}$,
we get the parameter $\beta'=3.996\approx4$, and we solve for $a_{4}$
and $b_{4}$ using the Eq. (\ref{eq:nNNSD_constants_for_Gaussian_ensembles}).
We achieve the analytical form presented in Eq. (\ref{eq:nNNSD_of_GOE})
{[}$P_{\mathrm{GOE}}^{2}(w_{2})${]}, which is also equivalent to
the corresponding NNSD of the standard GSE \cite{Mehta-Book}.
For the GUE class, using $\mathcal{H}_{\mathrm{GUE}}$, we get $\beta'=7.002\approx7$.
And using $\mathcal{H}_{\mathrm{GSE}}$, we find that the nNNSD of
the GSE class has a value $\beta'=13.015\approx13$. The
agreement of our computed effective Dyson index values for the nNNSD
of the three Gaussian ensembles with those deduced in Refs. \cite{santhanam_higher_order_ratio_nNNSD,Bhosale_higher_order_ratio_nNNSD,Wen-Jia_rao_Higher_order_level_spacing_nNNSD},
based on heuristic arguments and numerical studies, reasserts the
scaling relations proposed therein. We observe that, the values of
the effective Dyson indices $\beta'$, are \textit{universal}, i.e.,
they depend on the nature of the RMT ensembles, not on the kind of
spectral statistic studied. 

\begin{figure}[H]
\begin{centering}
\includegraphics[scale=0.5]{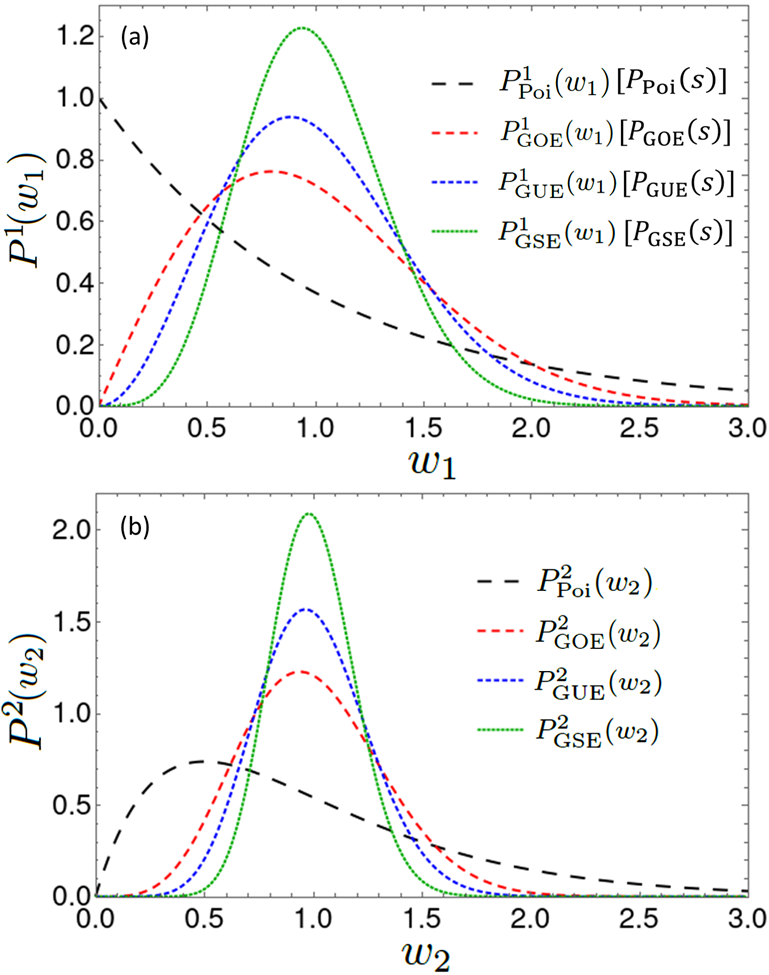}
\par\end{centering}
\caption{The comparison between the nearest-neighbor and next nearest-neighbor
spacing distributions; (a) presents the comparison between the Wigner-surmise
functional forms for the NNSD's of the Poissonian statistics, and
the three WD ensembles (see the expressions in the Table \ref{tab:NNSD_formulas_table}),
(b) presents the comparison between the Wigner-surmise-like functional
forms for the nNNSD's of the semi-Poissonian statistics, and the three
WD ensembles (see the expressions in the Table \ref{tab:nNNSD_formulas_table}).
\label{fig:nNNSD_combined}}
\end{figure}

With the normalization conditions taken care of, using Eq. \ref{eq:nNNSD_constants_for_Gaussian_ensembles},
we compile the values for the normalizing constants for the all symmetry
classes, and based on these values, compile the Wigner-surmise-like
formulas for the nNNSD distribution functions in Table \ref{tab:nNNSD_formulas_table}.
The nNNSD \textit{histograms} of the matrix models ($\mathcal{H}_{\mathrm{WD}}$),
along with the corresponding best-fitted $P_{\mathrm{WD}}^{2}(w_{2})$
distributions (from Table \ref{tab:nNNSD_formulas_table}) are plotted
in Figs. \ref{fig:nNNSD_GOE} (GOE), \ref{fig:nNNSD_GUE} (GUE), and
\ref{fig:nNNSD_GSE} (GSE), respectively. For comparison, the Wigner-surmise-like
functional forms for the nNNSD's (as summarized in Table \ref{tab:nNNSD_formulas_table})
for the integrable case (semi-Poissonian) along with all the Wigner-Dyson
ensembles, are plotted in Fig. \ref{fig:nNNSD_combined}(b). A similar
comparison for NNSD's is also plotted in Fig. \ref{fig:nNNSD_combined}(a).
We find that, just as for the NNSD, the nNNSD for the Wigner-Dyson
classes also show signs of increasing level-repulsion as one goes
from GOE to GUE to GSE, as depicted by the shifting of the peaks of
the distribution functions towards higher values of $w_{2}$, in that
order. Also the distribution function is zero at $w_{2}=0$ (degeneracies
not favored) and remains very small till about $w_{2}=0.25$, which
is somewhat larger than the thresholds for the NNSD case, as also
borne out by the much larger leading powers of $w_{2}$ in these expressions
(for small $w_{2}$, see Table \ref{tab:nNNSD_formulas_table}) compared
to those for the NNSD's (see Table \ref{tab:NNSD_formulas_table}). But more strikingly, even for the integrable case, the probability
is no longer large for $w_{2}=0$, but actually zero. However, since in the integrable case, the eigenlevels are uncorrelated,
this just implies that not more than two consecutive levels coincide
\cite{Wen-Jia_rao_Higher_order_level_spacing_nNNSD}. This is unlike
the correlated cases of random matrix ensembles, where level repulsion
arises due to the \textit{Vandermonde factor} \cite{Mehta-Book}, which involves the product of differences between
eigenvalues and leads to a vanishing probability density at zero spacing.

In the following section, we present exact analytical expressions
for the next nearest-neighbor spacing distributions based on $3\times3$
matrix models and derive a semi-analytical interpolation distribution
between Gaussian orthogonal and Gaussian unitary ensembles using three-dimensional
joint probability densities.

\begin{figure}[h]
\begin{centering}
\includegraphics[scale=0.5]{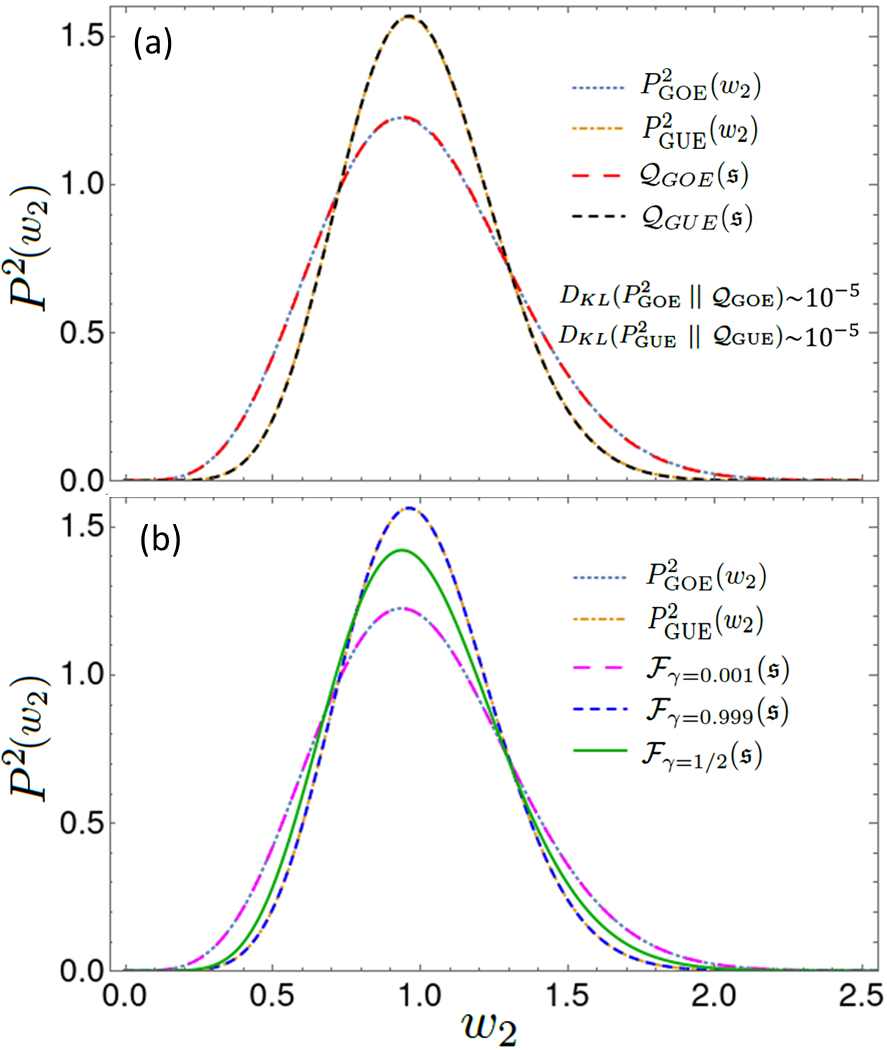}
\par\end{centering}
\caption{Comparison in nNNSD results. (a) shows comparison between the Wigner-surmise-like
distribution $P_{\mathrm{GOE}}^{2}(w_{2})$ {[}$P_{\mathrm{GUE}}^{2}(w_{2})${]}
and the analytical distribution $\mathcal{Q}_{\mathrm{GOE}}(\mathfrak{s})$
{[}$\mathcal{Q}_{\mathrm{GUE}}(\mathfrak{s})${]}, derived from the
three-dimensional JPDF of eigenvalues. They show excellent agreement,
as is also borne out by the very small ($\sim10^{-5}$) values of
the KL divergence measures {[}$D_{KL}(P_{\mathrm{GOE}}^{2}\mid\mid\mathcal{Q}_{\mathrm{GOE}})$
and $D_{KL}(P_{\mathrm{GUE}}^{2}\mid\mid\mathcal{Q}_{\mathrm{GUE}})${]}
between them. In (b), the semi-analytical distribution $\mathcal{F}_{\gamma}(\mathfrak{s})$
is plotted which beautifully interpolates between GOE and GUE limits
for the nNNSD. \label{fig:analytical_nNNSD_comparison}}
\end{figure}

\begin{widetext}

\section{Exact \texorpdfstring{\MakeLowercase{n}NNSD}{nNNSD} results for $3\times3$ Gaussian random
matrices\label{sec:Derivation-of-exact-nNNSD-results}}

Similar to Wigner-surmise formulae, which are based on $2\times2$
matrix models, one may employ $3\times3$ matrix models to deduce
analytical expressions for the nNNSD. These expressions serve as an
alternative to the Wigner-surmise results based on effective-$\beta$
($\beta'$) and are capable of effectively approximating large-dimensional
cases while also offering valuable additional insights. Notably, analytical
expressions for the nNNSD are already known for GOE and GUE due to
Porter and Kahn \cite{Porter_nNNSD_book_1965,Porter_nNNSD_GOE_1963,Porter_nNNSD_GUE_1963}.
Here, we present the scaled versions of these distributions which
ensure both the overall unit-normalization and the unit mean-nNNSD
conditions. For completeness, in the Appendix, the corresponding derivations
are included. Additionally, we tackle the GOE-to-GUE crossover using
the joint probability densities of three-dimensional Gaussian matrix
ensembles and obtain semi-analytical results which can be used to
capture the orthogonal-to-unitary crossover in\textcolor{red}{{} }physical
systems by relating the generic RMT crossover parameter to various
physical symmetry-breaking parameters, under different circumstances. 

\subsection{Exact Analytical Results of nNNSD for GOE and GUE \label{subsec:Exact-analytical-results-for_GOE_and_GUE}}

We present the probability density for the GOE {[}from Eq. (\ref{eq:normalized_analytical_GOE-Appendix})
in the Appendix{]},

\begin{equation}
\mathcal{Q}_{\mathrm{GOE}}(\mathfrak{s})=\frac{3^{4}\exp\left(-9\mathfrak{s}^{2}/\pi\right)\mathfrak{s}\left(6\mathfrak{s}-\exp\left(9\mathfrak{s}^{2}/4\pi\right)\left(2\pi-9\mathfrak{s}^{2}\right)\mathrm{erf}\left[\frac{3\mathfrak{s}}{2\sqrt{\pi}}\right]\right)}{4\pi^{2}},\label{eq:normalized_analytical_GOE}
\end{equation}
 where the $\mathfrak{s}$ is the scaled next nearest-neighbor spacing
variable and $\mathrm{erf}\left[z\right]=\frac{2}{\sqrt{\pi}}\int_{0}^{z}e^{-t^{2}}dt$
is the \textit{error function}. Also, the probability density for
GUE is given by {[}Eq. (\ref{eq:normalized_analytical_GUE-Appendix}){]}

\begin{equation}
\mathcal{Q}_{\mathrm{GUE}}(\mathfrak{s})=\frac{3^{11}\sqrt{3}\exp\left(-3^{5}\mathfrak{s}^{2}/2^{4}\pi\right)\mathfrak{s}^{2}\left(2^{3}3^{3}\mathfrak{s}\left(-2^{5}\pi+3^{4}\mathfrak{s}^{2}\right)+\sqrt{3}\exp\left(3^{5}\mathfrak{s}^{2}/2^{6}\pi\right)\left(2^{10}\pi^{2}-2^{6}3^{4}\pi\mathfrak{s}^{2}+3^{9}\mathfrak{s}^{4}\right)\mathrm{erf}\left[\frac{9}{8}\sqrt{\frac{3}{\pi}}\mathfrak{s}\right]\right)}{2^{20}\pi^{4}}.\label{eq:normalized_analytical_GUE}
\end{equation}
As mentioned in the Appendix, the scaled parameter $\mathfrak{s}$
is formally equivalent to the parameter $w_{2}$. So, to compare the
Wigner-surmise-like results, $P_{\mathrm{GOE}}^{2}(w_{2})$ and $P_{\mathrm{GUE}}^{2}(w_{2})$,
against the exact analytical distributions, $\mathcal{Q}_{\mathrm{GOE}}(\mathfrak{s})$
{[}$\equiv\mathcal{Q}_{\mathrm{GOE}}(w_{2})${]} and $\mathcal{Q}_{\mathrm{GUE}}(\mathfrak{s})$
{[}$\equiv\mathcal{Q}_{\mathrm{GUE}}(w_{2})${]}, respectively, we
perform their Taylor expansion about $w_{2}=1$, around which point
they possess maximum probability. For the GOE case, the expansion
up to third order gives, 
\begin{equation}
\lim_{w_{2}\rightarrow1}P_{\mathrm{GOE}}^{2}(w_{2})=1.206-0.635\,(w_{2}-1)-4.974\,(w_{2}-1)^{2}+4.288\,(w_{2}-1)^{3}+\cdots\label{eq:Taylor_expansion_GOE}
\end{equation}
and 
\begin{equation}
\lim_{w_{2}\rightarrow1}\mathcal{Q}_{\mathrm{GOE}}(w_{2})=1.201-0.650\,(w_{2}-1)-4.908\,(w_{2}-1)^{2}+4.361\,(w_{2}-1)^{3}+\cdots,\label{eq:Taylor_expansion_GOE_1}
\end{equation}
 and for the GUE, 
\begin{equation}
\lim_{w_{2}\rightarrow1}P_{\mathrm{GUE}}^{2}(w_{2})=1.551-0.801\,(w_{2}-1)-11.051\,(w_{2}-1)^{2}+9.398\,(w_{2}-1)^{3}+\cdots\label{eq:Taylor_expansion_GUE}
\end{equation}
and 
\begin{equation}
\lim_{w_{2}\rightarrow1}\mathcal{Q}_{\mathrm{GUE}}(w_{2})=1.546-0.817\,(w_{2}-1)-10.922\,(w_{2}-1)^{2}+9.527\,(w_{2}-1)^{3}+\cdots.\label{eq:Taylor_expansion_GUE_1}
\end{equation}
 By examining Eqs. (\ref{eq:Taylor_expansion_GOE}) and (\ref{eq:Taylor_expansion_GOE_1})
for the GOE and Eqs. (\ref{eq:Taylor_expansion_GUE}) and (\ref{eq:Taylor_expansion_GUE_1})
for the GUE, we observe that the maximum difference between the coefficients
in the Taylor expansion of the Wigner-surmise-like results and the
corresponding exact analytical results, ranges between $10^{-3}$
to $10^{-1}$. To add more credibility to our line of reasoning, in
Fig. \ref{fig:analytical_nNNSD_comparison}(a), we compare $P_{\mathrm{GOE}}^{2}(w_{2})$
{[}$P_{\mathrm{GUE}}^{2}(w_{2})${]} against the exact analytical
distribution $\mathcal{Q}_{\mathrm{GOE}}(\mathfrak{s})$ {[}$\mathcal{Q}_{\mathrm{GUE}}(\mathfrak{s})${]}.
We observe that, the corresponding distributions match very well,
which can also be confirmed from the extremely small values ($\sim10^{-5}$)
of their Kullback-Leibler (KL) divergences \cite{Debojyoti_paper_1_2022},
{[}$D_{KL}(P_{\mathrm{GOE}}^{2}\mid\mid\mathcal{Q}_{\mathrm{GOE}})$
and $D_{KL}(P_{\mathrm{GUE}}^{2}\mid\mid\mathcal{Q}_{\mathrm{GUE}})${]},
respectively, which measures the similarity of the two probability
distributions, as a whole. Next, we derive a semi-analytical interpolating
distribution for the GOE-to-GUE crossover.

\subsection{Derivation of Semi-analytical Result for nNNSD in GOE-to-GUE Crossover \label{subsec:Derivation-of-exact-analyrical-GOE-to-GUE_crossover}}

To start with, we consider the well known form of the Pandey-Mehta
crossover matrix model \cite{PandeyMehta1983,MehtaPandey1983,Lenz_Haake_crossover_matrix,Ayana's_PhysRevE}:
\begin{equation}
\mathcal{H}=\sqrt{1-\gamma^{2}}\,\mathcal{H}_{0}+\gamma\,\mathcal{H}_{1},\label{eq:crossover_matrix_model_1}
\end{equation}
 where the crossover parameter $\gamma$ is varied from 0 to 1 and
leading to an interpolation between the two RMT ensembles having characteristic
Hamiltonians $\mathcal{H}_{0}$ and $\mathcal{H}_{1}$, respectively.
For our purpose, we consider the GOE-to-GUE crossover matrix model:
\begin{equation}
\mathcal{H}_{\mathrm{GOE-GUE}}=\sqrt{1-\gamma^{2}}\,\mathcal{H}_{\mathrm{GOE}}+\gamma\,\mathcal{H}_{\mathrm{GUE}},\label{eq:crossover_matrix_model_2}
\end{equation}
 which interpolates between the GOE and GUE symmetry classes. 

Now, from the Refs. \cite{PandeyMehta1983,MehtaPandey1983,Mehta-Book,Ayana's_PhysRevE},
we get the analytical expression of the joint probability distribution
function for the GOE-to-GUE crossover, $p_{\gamma}(x_{1,}x_{2},x_{3})$,
which uses the three-dimensional crossover matrix model $\mathcal{H}_{\mathrm{GOE-GUE}}$
{[}Eq. (\ref{eq:crossover_matrix_model_2}){]}, and given by 
\begin{equation}
p_{\gamma}(x_{1,}x_{2},x_{3})=\frac{1}{48\sqrt{2}\pi(1-\gamma^{2})^{3/2}}\left[f(x_{1}-x_{2})+f(x_{2}-x_{3})-f(x_{1}-x_{3})\right](x_{1}-x_{2})(x_{2}-x_{3})(x_{1}-x_{3})e^{-\frac{1}{4}(x_{1}^{2}+x_{2}^{2}+x_{3}^{2})},\label{eq:analytical_GOE-GUE_JPDF}
\end{equation}
 where $f(u)=\mathrm{erf}\left[\left(\frac{1-\gamma^{2}}{8\gamma^{2}}\right)^{1/2}u\right]$.
Now, using Eq. (\ref{eq:analytical_general_ensemble_probability_density}),
we get 
\begin{equation}
F_{\gamma}(\mathfrak{s})=6\int_{-\infty}^{\infty}dx_{2}\int_{-\infty}^{x_{2}}dx\int_{x_{2}}^{\infty}dy\;p_{\gamma}(x_{2}-x,x_{2},x_{2}+y)\delta\left[\mathfrak{s}-(x+y)\right],\label{eq:analytical_GOE-GUE_JPDF_1}
\end{equation}
 and after integrating over $x_{2}$, we are left with 
\begin{equation}
F_{\gamma}(\mathfrak{s})=\frac{1}{4\sqrt{6\pi}(1-\gamma^{2})^{3/2}}\int_{0}^{\infty}dx\int_{0}^{\infty}dy\;e^{-\frac{(x^{2}+xy+y^{2})}{6}}xy(x+y)\left[f(x)+f(y)-f(x+y)\right]\delta\left[\mathfrak{s}-(x+y)\right].\label{eq:analytical_GOE-GUE_JPDF_2}
\end{equation}
 After performing the integration over $y$, we get the semi-analytical
distribution for GOE-to-GUE interpolation as 
\begin{equation}
F_{\gamma}(\mathfrak{s})=\frac{1}{4\sqrt{6\pi}(1-\gamma^{2})^{3/2}}\int_{0}^{\mathfrak{s}}dx\;e^{-\frac{(-\mathfrak{s}^{2}+\mathfrak{s}x-x^{2})}{6}}\mathfrak{s}(\mathfrak{s}-x)x\left[f(x)+f(\mathfrak{s}-x)-f(\mathfrak{s})\right].\label{eq:analytical_GOE-GUE_JPDF_3}
\end{equation}
 Unfortunately, the above integration cannot be
evaluated in a closed analytical form, therefore, we adopt a \textit{semi-analytical}
approach. We consider the integrand in the above equation as the density
\begin{equation}
f_{\gamma}(\mathfrak{s},x)=\frac{1}{4\sqrt{6\pi}(1-\gamma^{2})^{3/2}}e^{-\frac{(-\mathfrak{s}^{2}+\mathfrak{s}x-x^{2})}{6}}\mathfrak{s}(\mathfrak{s}-x)x\left[f(x)+f(\mathfrak{s}-x)-f(\mathfrak{s})\right],\label{eq:analytical_GOE-GUE_JPDF_4}
\end{equation}
 and then numerically evaluate the integral for different values of
the crossover parameter $\gamma$. Next, we also need to normalize
$F_{\gamma}(\mathfrak{s})$ for each values of $\gamma$. For example,
for the $\gamma=1/2$ case, $\int_{0}^{\infty}d\mathfrak{s}\int_{0}^{\mathfrak{s}}f_{\gamma=1/2}(\mathfrak{s},x)dx=1$
and $\int_{0}^{\infty}\mathfrak{s}d\mathfrak{s}\int_{0}^{\mathfrak{s}}f_{\gamma=1/2}(\mathfrak{s},x)dx=4.52873=\kappa$.
Now, we get the corresponding normalized nNNSD function with unit
mean density, as $\mathcal{F}_{\gamma=1/2}(\mathfrak{s})=\kappa F_{\gamma=1/2}(\kappa\mathfrak{s})$.
Similarly, we get $\mathcal{F}_{\gamma}(\mathfrak{s})$ for various
values of the crossover parameter $\gamma$.

In Fig. \ref{fig:analytical_nNNSD_comparison}(b), we plot the semi-analytical
interpolating distribution $\mathcal{F}_{\gamma}(\mathfrak{s})$ for
the GOE-to-GUE crossover. The limiting distributions, $\mathcal{F}_{\gamma=0.001}(\mathfrak{s})$
($\gamma\rightarrow0$) and $\mathcal{F}_{\gamma=0.999}(\mathfrak{s})$
($\gamma\rightarrow1$), match well with $P_{\mathrm{GOE}}^{2}(w_{2})$
and $P_{\mathrm{GUE}}^{2}(w_{2})$, respectively. We also plot one
of the intermediate cases, $\mathcal{F}_{\gamma=1/2}(\mathfrak{s})$,
which lies between the limiting distributions. In the next section,
we study the robustness of all the nNNSD results obtained both in
this section and in the previous Sec. \ref{sec:METHODOLOGY-and-CALCULATIONS},
by comparing them against the corresponding distributions in a single-body
and two many-body quantum chaotic systems. 
\end{widetext}

\section{Study of \texorpdfstring{\MakeLowercase{n}NNSD}{nNNSD} in PHYSICAL Systems\label{sec:Study-of-nNNSD-in-Quantum-Chaotic-Systems}}

In view of the discussions in the previous sections, and applications of NNSD universality to various physical systems~\cite{Modak's-paper_2014,Avishai-Richert-PRB,Collura_Bose_Hubbard_2,Chavda-Kota-PLA-1,Iyer_Oganesyan_quasi-periodic_system,Debojyoti_paper_1_2022,Debojyoti_paper_2_2023},
based on the standard Berry-Tabor \cite{Berry-Tabor_conjecture_paper_1977.0140} and BGS \cite{BGS-conjecture}
conjectures, it is expected that the emergence of the nNNSD Statistics as shown by the Wigner-surmise-like results of
Table \ref{tab:nNNSD_formulas_table} and also embodied by the \textit{exact}
Eqs. (\ref{eq:normalized_analytical_GOE}) and (\ref{eq:normalized_analytical_GUE}), should also
be indicative of whether a model physical system is integrable or ergodic (chaotic), thus generalising the results of the aforesaid conjectures.
%Though some demonstrations of applying the next nearest-neighbor ratio distribution to physical systems exist  \cite{santhanam_higher_order_ratio_nNNSD,Chavda_Distribution_of_Higher_Order_Spacing_Ratios,Bhosale_higher_order_ratio_nNNSD}
%and they agree reasonably well with the RMT predictions, a systematic
%and unified demonstration of the predicted nNNSD and crossover formulas
%to a wide variety of physical systems does not exist to our knowledge.
To test this and check the robustness of the nNNSD universality and their
extent of agreement with the spectral statistics of physical systems,
we now proceed to study several physical systems and compare their
eigenvalue statistics with the results derived above. Specifically
we study four diverse systems in all. The first two are non-interacting
single-body models with well-known classically integrable and chaotic
limits. These are the Quantum Kicked Rotor (QKR) and the Quantum Quarter
Sinai Billiard (QQSB). The latter two systems are two different models
of interacting spin-$\frac{1}{2}$'s on a 1D lattice, which have no
classical counterpart whatsoever. 

\begin{figure*}
\begin{centering}
\includegraphics[scale=0.55]{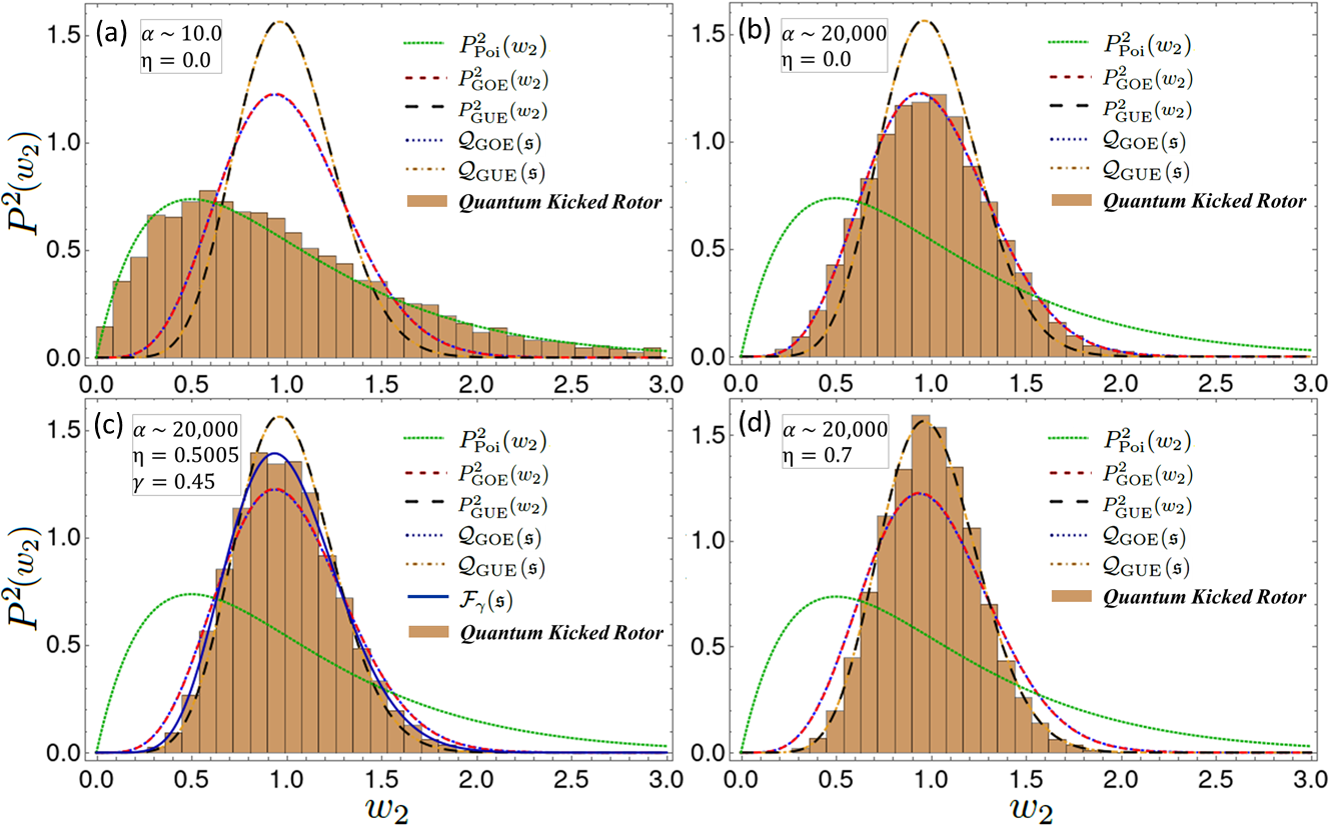}
\par\end{centering}
\caption{The nNNSD's of eigenangles for the single-body QKR system. (a) Shows
that the nNNSD of the QKR system faithfully follows the semi-Poissonian
distribution {[}$P_{\mathrm{Poi}}^{2}(w_{2})${]} for $\alpha\sim10$.
In (b) and (d), we plot the nNNSD, for the QKR system, in the GOE
limit ($\eta=0$) and GUE limit ($\eta=0.7$), respectively. They
follow the corresponding Wigner-surmise-like expressions, $P_{\mathrm{GOE}}^{2}(w_{2})$
and $P_{\mathrm{GUE}}^{2}(w_{2})$, as also by the exact analytical
distributions, $\mathcal{Q}_{\mathrm{GOE}}(\mathfrak{s})$ and $\mathcal{Q}_{\mathrm{GUE}}(\mathfrak{s})$,
as expected. In (c), the intermediate distribution ($\eta=0.5005$)
, in the crossover region, is also seen to be perfectly reproduced
by our derived semi-analytical distribution $\mathcal{F}_{\gamma}(\mathfrak{s})$
with $\gamma=0.45$. \label{fig:floquet_QKR_nNNSD}}
\end{figure*}

\subsection{Non-interacting Quantum Systems with Well-known Classically Chaotic
Limits \label{subsec:Quantum-Systems-with-classically_chaotic_limit}}

\subsubsection{The Quantum Kicked Rotor (QKR) \label{sec:Quantum-Kicked-Rotor}}

It is well-known that in 1D systems which are conservative, no chaos
is possible as there is one degree of freedom and one conserved quantity
(energy). So the only way to overcome this is to break the energy
conservation by introducing a time-dependent term with a tunable amplitude
so that the integrability may be broken gradually. In addition, if
the time-dependent term is periodic in time, although energy is not
strictly conserved, a discretized \textit{quasi-energy} is conserved,
much akin to the conservation of lattice momentum in spatially periodic
crystals. The kicked rotor system is a well-known prototype of the
above kind, and a single-body chaotic model that has been investigated
in both classical \cite{Chirikov_CERN_report_classical_QKR_1969,Lichtenberg_book_classical_QKR_2013}
and quantum regimes \cite{Izrailev_first_paper_QKR,Casati_QKR,Izrailev_QKR_2,Santhanam_review_QKR_2022,Izrailev_second_paper_QKR_1988}.
The NNSD of the quantum kicked rotor system follows the universal
Wigner-Dyson distributions in the strongly chaotic limit \cite{Izrailev_first_paper_QKR,Izrailev_QKR_2,Pandey_QKR,Izrailev_second_paper_QKR_1988}.
In this section, we investigate the nNNSD of the QKR system.

In a kicked rotor, a particle is confined to a ring (periodic boundary
conditions) and exposed to periodic kicks that vary with its angular
position ($\theta$). In order to also simulate the GUE regime via
the breaking of \textit{time-reversal symmetry}, we have introduced
an effective magnetic field via a minimal-coupling like term ($\eta$)
in the kinetic energy term. The Hamiltonian of the kicked rotor system
is given by \cite{Casati_QKR,Izrailev_QKR_2,Santhanam_review_QKR_2022,Ayana's_PhysRevE,Pandey_Jaiswal_QKR_2019}
\begin{equation}
H_{KR}=\frac{(p+\eta)^{2}}{2}+\alpha\cos(\theta+\theta_{0})\sum_{m=-\infty}^{\infty}\delta(t-m),\label{eq:Kicked_rotor_Hamiltonian}
\end{equation}
 where $p$ represents the angular momentum operator and $\alpha$
is the \textit{kicking} (or \textit{stochasticity}) parameter. The
quantity $\theta_{0}$ is the \textit{parity} breaking parameter and
$\eta$ ($0\leq\eta<1$) is the \textit{time-reversal} symmetry breaking
parameter. The Dirac comb {[}$\sum_{m=-\infty}^{\infty}\delta(t-m)${]}
is employed to deliver periodic kicks to the particle at integer time
intervals ($t=m$). Using Floquet's theorem \cite{Haake-Chaos_book},
the unitary \textit{time evolution} operator $U=BG$, guides the quantum
dynamics of the QKR system, where, $B(\alpha)=\exp\left[-\frac{i}{\hbar}\alpha\cos(\theta+\theta_{0})\right]$
is a function of the \textit{kicking} parameter $\alpha$, and $G=\exp\left[-\frac{i}{2\hbar}(p+\eta)^{2}\right]$
\cite{Pandey_QKR,Pandey_Jaiswal_QKR_2019,Pandey_Puri_QKR_2020}.
Under the \textit{torus} boundary condition, the position ($\theta$)
and momentum ($p$) operators give discrete eigenvalues and $U$ is
reduced to a finite $\mathcal{N}\times\mathcal{N}$ dimensional matrix
form. The position representations of $B$ and $G$ are \cite{Pandey_Jaiswal_QKR_2019,Pandey_Puri_QKR_2020,Ayana's_PhysRevE,Izrailev_first_paper_QKR,Izrailev_second_paper_QKR_1988}
\begin{equation}
B_{kl}=\exp\left[-i\alpha\cos\left(\frac{2\pi k}{\mathcal{N}}+\theta_{0}\right)\right]\delta_{kl}\label{eq:Expression_of_B_kl_in_QKR}
\end{equation}
 and 
\begin{equation}
G_{kl}=\frac{1}{\mathcal{N}}\sum_{j=-\frac{(\mathcal{N}-1)}{2}}^{\frac{(\mathcal{N}-1)}{2}}\exp\left[-i\left(\frac{j^{2}}{2}-\eta j-2\pi j\frac{(k-l)}{\mathcal{N}}\right)\right],\label{eq:Expression_of_G_kl_in_QKR}
\end{equation}
 where $k,l=-(\mathcal{N}-1)/2,-(\mathcal{N}-1)/2+1,\ldots,(\mathcal{N}-1)/2$
and we set $\hbar=1$. From this, it is clear that the indices $(k,l)$
are equivalent to the magnetic quantum number $m_{\mathcal{J}}$ (where
$\mathcal{J}$ is the value of the angular momentum involved) and
${\cal N}=(2\mathcal{J}+1)$ is the degeneracy of that level. Evidently
for large values of ${\cal N}$ the system should exhibit semi-classical
behavior. In the position basis, the $kl$-th element of the unitary
matrix of the time evolution operator becomes \cite{Pandey_Jaiswal_QKR_2019,Pandey_Puri_QKR_2020,Ayana's_PhysRevE,Izrailev_first_paper_QKR,Izrailev_second_paper_QKR_1988}
\begin{align} \nonumber U_{kl}=&\frac{1}{\mathcal{N}}\exp\left[-i\alpha\cos\left(\frac{2\pi k}{\mathcal{N}}+\theta_{0}\right)\right] \\ &\times\sum_{j=-\frac{(\mathcal{N}-1)}{2}}^{\frac{(\mathcal{N}-1)}{2}}\exp\left[-i\left(\frac{j^{2}}{2}-\eta j-2\pi j\frac{(k-l)}{\mathcal{N}}\right)\right].\label{eq:Expression_of_U_kl_in_QKR} \end{align}

The eigenvalues of the unitary matrix $U$ are phase factors of the
form $e^{i\varphi_{j}}$ and they lie on a unit circle. The ordered
sequence of the \textit{eigenangles}, $0\leq\varphi_{1}<\varphi_{2}\cdots<\varphi_{N_{1}}<2\pi$,
are considered to study the level spacing distributions for the QKR
system. Using the unfolded eigenangles {[}$\tilde{\varphi}_{j}=\varphi_{j}(\mathcal{N}/2\pi)${]},
we can define the next-nearest neighbor spacings $(w_{2})_{j}=(\tilde{\varphi}_{j+1}-\tilde{\varphi}_{j})/2$.
Even though the classical kicked rotor system exhibits chaotic motion
when subjected to a relatively low kicking parameter ($\alpha\sim10$),
the NNSD of the QKR system still adheres to Poissonian statistics.
This can be attributed to the presence of localization effects \cite{Pandey_QKR,Pragya_Shukla_1997,Casati_QKR_2,Mirlin_Scaling_properties_localization_random_band_matrices,Santhanam_review_QKR_2022,Pandey_Jaiswal_QKR_2019}.
When the parity symmetry is fully broken ($\theta_{0}=\pi/2\mathcal{N}$),
and the kicking parameter $\alpha$ is very large ($\alpha^{2}\gg\mathcal{N}$),
the time-reversal symmetry of the QKR system is \textit{preserved}
(\textit{broken}) for $\eta=0$ ($\eta\neq0$) and the level correlation
statistics follows the universal GOE (GUE) distribution \cite{Pandey_Jaiswal_QKR_2019,Pandey_Puri_QKR_2020,Pandey_QKR}. 

In Fig. \ref{fig:floquet_QKR_nNNSD}(a), we examine the nNNSD in the
QKR system with $\alpha\sim10$ (integrable limit) and for an ensemble
of $\mathcal{M}=20$ matrices\footnote{For the integrable case, the parameter $\alpha$ is varied from 11
to 30 and for the chaotic case, it is varied from 20,001 to 20,020,
to construct the matrix ensembles.\label{fn:Ensemble_for_QKR}}, for $\mathcal{N}=501$ and observe its semi-Poissonian behavior.
In Fig. \ref{fig:floquet_QKR_nNNSD}(b)-(d), we present the nNNSD
of the QKR system in the quantum chaotic regime with $\alpha\sim2\times10^{4}$
($\mathcal{M}=20$\ref{fn:Ensemble_for_QKR} and $\mathcal{N}=501$).
Fig. \ref{fig:floquet_QKR_nNNSD}(b) represents the time-reversal
symmetry invariant QKR system with $\eta=0$ \cite{Pandey_Jaiswal_QKR_2019,Pandey_Puri_QKR_2020,Pandey_QKR}
and the plotted nNNSD follows the Wigner-surmise-like distribution
$P_{\mathrm{GOE}}^{2}(w_{2})$, as one might expect. We also notice
that, the \textit{exact} distribution for GOE, $\mathcal{Q}_{\mathrm{GOE}}(\mathfrak{s})$,
matches well with the spectral correlation statistics of the quantum
chaotic system, in this limit. Similarly, for a time-reversal symmetry
broken QKR system ($\eta=0.7$) \cite{Pandey_Jaiswal_QKR_2019,Pandey_Puri_QKR_2020},
we plot the nNNSD in Fig. \ref{fig:floquet_QKR_nNNSD}(d), and it
follows the distribution $P_{\mathrm{GUE}}^{2}(w_{2})$, quite closely.
Also, the time-reversal symmetry broken QKR system seems to follow
the \textit{exact} analytical distribution $\mathcal{Q}_{\mathrm{GUE}}(\mathfrak{s})$
for the nNNSD (GUE regime), as one would expect. In Fig. \ref{fig:floquet_QKR_nNNSD}(c),
we plot an intermediate (partially symmetry-broken) case with $\eta=0.5005$,
which can be perfectly mapped by our derived semi-analytical distribution
$\mathcal{F}_{\gamma}(\mathfrak{s})$ with $\gamma=0.45$.

\begin{figure}[h]
\begin{centering}
\includegraphics[scale=0.5]{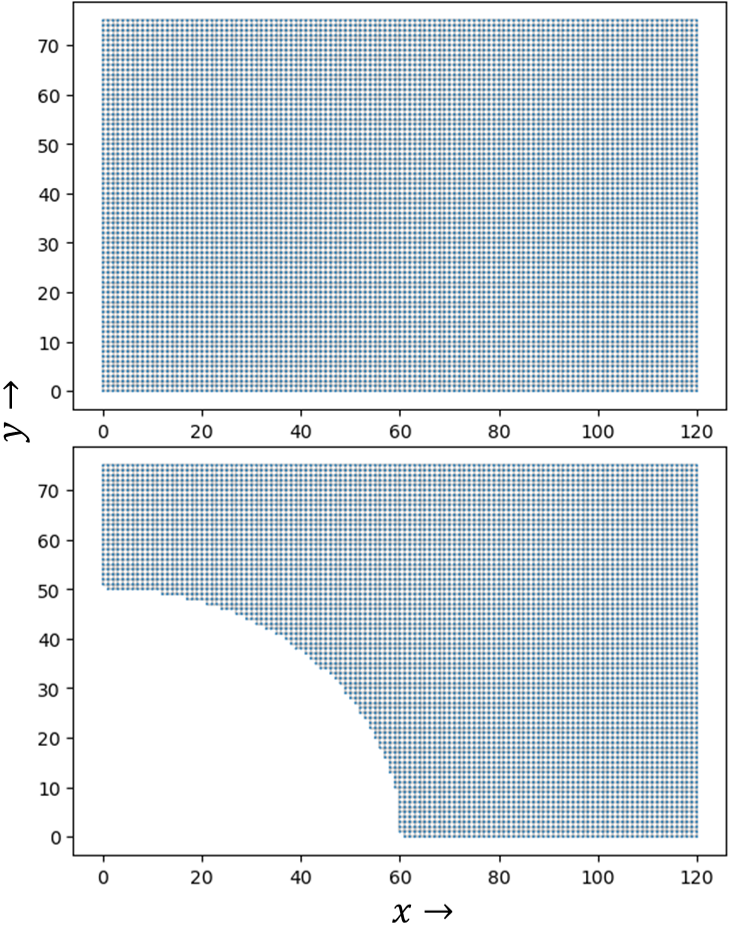}
\par\end{centering}
\caption{Top: the Kwant-generated schematic of an integrable rectangular billiard
with dimension $120\times75$ and the corresponding tight-binding
matrix dimension is $9196$. Bottom: the schematic of a quarter Sinai
billiard system, represented in a $120\times75$ rectangular lattice
using Kwant by removing a \textit{quarter} elliptical shape from the
lower left corner. The elliptical shape follows the equation $x^{2}/60^{2}+y^{2}/50^{2}=1$.
The dimension of the corresponding tight-binding matrix is $6789$.
\label{fig:Sinai_Billiard_representation}}
\end{figure}

\subsubsection{The Quantum Quarter Sinai Billiard (QQSB) \label{subsec:Quarter-Sinai-Billiard}}

Next we proceed to a family of non-interacting systems in 2D, namely
the Quantum Billiards \cite{billiard_1999,Bunimovich_billiard_1974,Bunimovich_billiard_1979,Cuevas_quantum_billiard_1996,Dietz_Dirac_Billiard_2016,Frahm_billiard_1995,Huang_level_spacing_statistics_graphene_billiards,Kota_Embedded_Random_Matrix_Ensembles_billiards,sinai_billiard_1963,Sinai_billiard_1970,Yu_graphene_billiard_2016,Santosh-Rohit2020,BGS-conjecture,Izrailev_first_paper_QKR},
which have classically integrable or chaotic counterparts. In two
dimensions, integrability requires the conservation of one more quantity,
in addition to energy, so that even conservative systems may exhibit
chaos, under suitable conditions. For example, in a circular billiard,
both energy and the component of angular momentum along the perpendicular
axis through the center are conserved, and it shows no chaos. But
for a Sinai billiard \cite{sinai_billiard_1963,Sinai_billiard_1970,Santosh-Rohit2020}
(square or rectangular shape, with a circular or elliptic hole cut
out at the center), or for a Bunimovich stadium \cite{Bunimovich_billiard_1974,Bunimovich_billiard_1979,Santosh-Rohit2020},
angular momentum is no longer conserved and a chaotic regime is possible.

Thus, the quantum billiard systems are potential single-body quantum-chaotic
systems which have classically chaotic counterparts \cite{billiard_1999,Bunimovich_billiard_1974,Bunimovich_billiard_1979,Cuevas_quantum_billiard_1996,Dietz_Dirac_Billiard_2016,Frahm_billiard_1995,Huang_level_spacing_statistics_graphene_billiards,Kota_Embedded_Random_Matrix_Ensembles_billiards,sinai_billiard_1963,Sinai_billiard_1970,Yu_graphene_billiard_2016,Santosh-Rohit2020,BGS-conjecture,Izrailev_first_paper_QKR}.
The nearest-neighbor spectral fluctuation statistics of quantum chaotic
billiards has been widely explored by RMT studies, pioneered by the
ground-breaking work of Bohigas, Giannoni and Schmit in the famous
BGS Conjecture paper \cite{BGS-conjecture}. One such system is
the \textit{quarter Sinai billiard} \cite{sinai_billiard_1963,Sinai_billiard_1970,Santosh-Rohit2020}.
In this study, we employ Kwant \cite{Groth_Kwant_quantum_transport_2014},
an open source computational tool, to analyze the nNNSD's of the discretized
version of the tight-binding (TB) Hamiltonian in the two-dimensional
quarter Sinai billiard system, operating in the chaotic regime. The
discretized tight-binding model Hamiltonian is given by \cite{Groth_Kwant_quantum_transport_2014,Santosh-Rohit2020}
\begin{align} \nonumber H_{TB}&=\sum_{ij}[\left(V(ak,al)+4t\right)\left|k,l\right\rangle \left\langle k,l\right|-t(\left|k+1,l\right\rangle \left\langle k,l\right| \\ &+\left|k,l\right\rangle \left\langle k+1,l\right|+\left|k,l+1\right\rangle \left\langle k,l\right|+\left|k,l\right\rangle \left\langle k,l+1\right|)], \label{eq:Sinai_billiard_TB_Hamiltonian} \end{align}
where $t=\hbar^{2}/(2ma^{2})$ is the hopping parameter with $a$
as the lattice constant and $m$ as the mass of the particle. $\left|k,l\right\rangle $
($\equiv\left|ak,al\right\rangle =\left|x,y\right\rangle $) represent
the discretized positional states corresponding to the potential $V(x,y)$. 

\begin{figure*}
\begin{centering}
\includegraphics[scale=0.55]{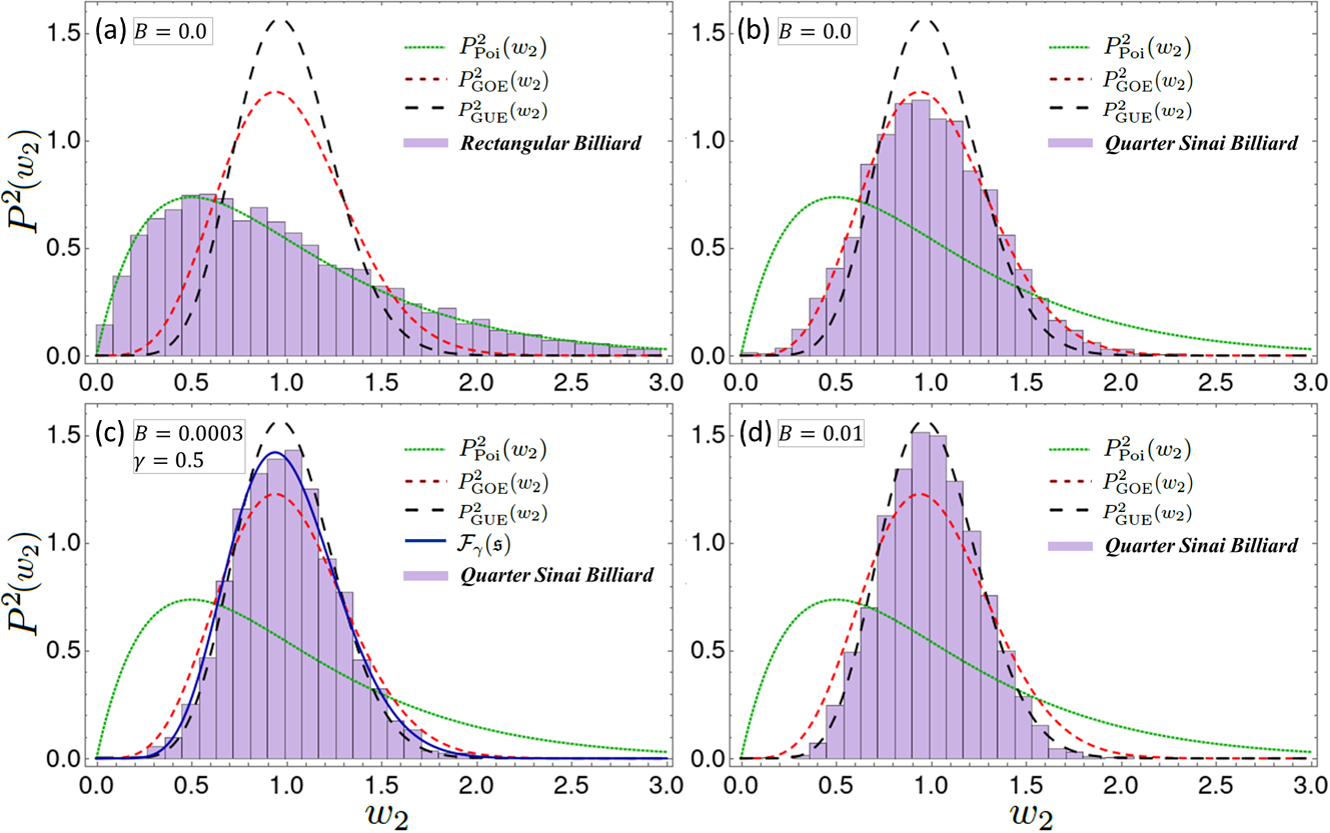}
\par\end{centering}
\caption{The nNNSD's for the single-body (a) integrable rectangular billiard
and (b)-(d) chaotic quarter Sinai billiard systems. (a) Shows the
semi-Poissonian statistics for the rectangular billiard with ($B=0$
with the tight-binding matrix dimension $9196$). In (b) and (d),
we plot the nNNSD for the quarter Sinai billiard system (with the
tight-binding matrix dimension $6789$) in the GOE limit ($B=0$)
and GUE limit ($B=0.01$), respectively. In (c), the intermediate
distribution ($B=0.0003$), in the crossover region, is perfectly
mapped by our derived semi-analytical distribution $\mathcal{F}_{\gamma}(\mathfrak{s})$,
with $\gamma=0.5$. \label{fig:Sinai_Billiard_nNNSD}}
\end{figure*}

If one applies a time-reversal symmetry breaking, uniform magnetic
field to the system, it modifies the hopping term $t$ by introducing
an additional phase factor, $\exp[-\frac{ie}{\hbar c}\int_{\mathbf{r}_{1}}^{\mathbf{r}_{2}}d\mathbf{r}\cdot\mathbf{A}(\mathbf{r})]$,
known as the Peierls or the Aharonov-Bohm phase, where $\mathbf{A}(\mathbf{r})$
is the corresponding vector potential. We use the Peierls substitution
scheme \cite{Santosh-Rohit2020,Peierls_Kwant_substitution_scheme,Hofstadter_Peierls_Kwant_substitution_scheme,Groth_Kwant_quantum_transport_2014}
in the Kwant code to implement the hopping between two lattice points,
$\mathbf{r}_{1}$ and $\mathbf{r}_{2}$. For the two-dimensional billiard,
a choice of gauge, $\mathbf{A}\equiv(A_{x},A_{y},A_{z})\equiv(-By,0,0)$,
ensures the constant and uniform magnetic field $B$ in the $z$-direction
(perpendicular to the plane of the lattice), and the phase factor
becomes $\exp(iaBy)$ {[}with $e/(\hbar c)=1${]} \cite{Santosh-Rohit2020,Peierls_Kwant_substitution_scheme}.
So, the hopping parameter $t$ along the $x$-direction now becomes
$t*\exp(iaBy)$, whereas it remains unchanged along the $y$-direction.

We consider the \textit{particle in a box} approximation and set $V(ak,al)=0$
in it. Following the Ref. \cite{Santosh-Rohit2020}, we use the
Kwant code \cite{Groth_Kwant_quantum_transport_2014} and implement
the quarter Sinai billiard system on a $120\times75$ rectangular
lattice by removing a \textit{quarter} elliptical portion from the
left lower corner of it. The elliptical shape follows the equation
$x^{2}/60^{2}+y^{2}/50^{2}=1$. The dimension of the corresponding
tight-binding matrix is $6789$ and we set $a=t=1$ throughout this
calculation. In Fig. \ref{fig:Sinai_Billiard_representation} (\textit{bottom}),
we present the Kwant generated image for this two-dimensional quarter
Sinai billiard system. The pure \textit{rectangular billiard} is known
to be classically integrable. To compare the spectral correlations
of the quantum rectangular billiard system against the chaotic quarter
Sinai billiard system, we also carry out our Kwant calculations for
this system on a $120\times75$ rectangular lattice and present the
corresponding schematic in Fig. \ref{fig:Sinai_Billiard_representation}
(\textit{top}). The dimension of the corresponding tight-binding matrix
is $9196$.

For $B=0$ ($B\neq0$), the time-reversal symmetry of the system is
preserved (broken), the spectral correlations of the quarter Sinai
billiard system are expected to follow the GOE (GUE) statistics \cite{Santosh-Rohit2020}.
In Fig. \ref{fig:Sinai_Billiard_nNNSD}(b)-(d), we plot the nNNSD
of this single-body chaotic billiard system. Fig. \ref{fig:Sinai_Billiard_nNNSD}(b)
represents the $B=0$ scenario, which follows the distribution $P_{\mathrm{GOE}}^{2}(w_{2})$.
To avoid the emergence of Landau levels and the resulting deviations
from RMT behavior \cite{Santosh-Rohit2020}, we use a small magnetic
field, $B=0.01$, and plot the corresponding nNNSD for the billiard
system in Fig. \ref{fig:Sinai_Billiard_nNNSD}(d), which matches well
with the distribution $P_{\mathrm{GUE}}^{2}(w_{2})$. In Fig. \ref{fig:Sinai_Billiard_nNNSD}(c),
we present an intermediate case, $B=0.0003$, which is well described
by our derived semi-analytical distribution $\mathcal{F}_{\gamma}(\mathfrak{s})$
with $\gamma=0.5$. To compare with the spectral correlation statistics
of the integrable system, we also calculate the nNNSD of the rectangular
billiard system and plot its semi-Poissonian {[}$P_{\mathrm{Poi}}^{2}(w_{2})${]}
statistical behavior in Fig. \ref{fig:Sinai_Billiard_nNNSD}(a).

\subsection{Chaos in Quantum Many-body Systems with No Classical Analogue: Disordered
Spin-chain Models\label{sec:Disordered-spin-chain-models}}

In this Section, we perform calculations on two disordered spin-chain
models, which we refer to as model-I and model-II, already introduced
in our earlier works \cite{Debojyoti_paper_1_2022,Debojyoti_paper_2_2023}.
The competition between different Hamiltonian terms can break or preserve
various unitary and antiunitary symmetries and therefore determine
the spectral correlation statistics of the Hamiltonian matrices of
these many-body correlated systems, as discussed at length in References
\cite{Debojyoti_paper_1_2022,Debojyoti_paper_2_2023}, in the context
of NNSD and RD calculations in various regimes. In this section, we
study the much less studied nNNSD's of these systems, and compare
them with the exact analytical results obtained in the previous sections.

\subsubsection{Model-I}

Our model-I is a one-dimensional half-filled spin-$\frac{1}{2}$ Hamiltonian,
given by \cite{Debojyoti_paper_1_2022}, \begin{align} 
\nonumber H_1 & =H_{h}+H_{r}+H_{c}\\ \nonumber & =\sum_{j=1}^{N-1}J\boldsymbol{\mathrm{S}}_{j}\cdot\boldsymbol{\mathrm{S}}_{j+1}+\sum_{j=1}^{N}h_{j}\mathrm{S^{z}}_{j}\\ & +\sum_{j=1}^{N-2}J_{t}\mathbf{S}_{j}\cdot[\mathbf{S}_{j+1}\times\mathbf{S}_{j+2}],
\label{eq:model-I_Hamiltonian} 
\end{align} which has $N$ sites with one spin per lattice site. In this model,
the isotropic Heisenberg term ($H_{h}$) \cite{Heisenberg_original_paper,Bloch_Heisenberg_model_paper_1,Bloch_Heisenberg_model_paper_2}
is coupled to a randomly distributed and spatially inhomogeneous magnetic
field by the random Zeeman term $H_{r}$. It also includes the 3-site
scalar spin-chirality term $H_{c}$. This type of spin correlation
term, may appear when a circularly polarized laser is used to probe
Mott insulators \cite{Diptiman's-paper,spin_chirality_Kitamura,Avishai-Richert-PRB,Modak's-paper_2014,Debojyoti_paper_1_2022}.
Here $\boldsymbol{\mathrm{S}}_{j}$ is the vector spin operator at
site $j$, and $J$ is the nearest-neighbor exchange interaction.
Disorder is introduced in the Zeeman term via the parameter $h_{j}$,
a random magnetic field along $\mathrm{z}$-direction at site $j$,
drawn from a Gaussian distribution having zero mean and variance $h^{2}$.
$J_{t}$ is the coupling constant, which appears in the three-site
scalar spin-chirality term $H_{c}$.

For constructing the Hamiltonian matrix, we consider a direct product basis of single-site  spin-1/2's . Each site can occupy either of an up-spin ($\mathrm{m^{z}}=1/2\equiv$ $\uparrow$) or a down-spin ($\mathrm{m^{z}}=-1/2\equiv$ $\downarrow$), with $\mathrm{m^{z}}$ being the eigenvalue of the $\mathrm{z}$-component of the spin operator $\mathrm{S^{z}}$. There exists $2^{N}$ number of basis states ($\left|\mathrm{m_{1}^{z}m_{2}^{z}m_{3}^{z}\mathbf{....}m_{\mathit{N}}^{z}}\right\rangle $) for an $N$-site system. $\mathrm{S^{z}}=\mathrm{\sum_{j=1}^{\mathit{N}}}\mathrm{S}_{j}^{\mathrm{z}}$ is the $\mathrm{z}$-component of the total spin and we find that $\left[\mathrm{S^{z}},H_1\right]=0$. $H_1$ consists of irreducible blocks corresponding to different $\mathrm{S^{z}}$ subsectors. As a result, the eigenvalues of these subsectors are not correlated, whereas the eigenvalues within a given sector are correlated. For RMT analysis consideration, we need to work with one such block. In our calculations, we consider an \emph{even} $N$ and the $\mathrm{S^{z}}=0$ subsector (Hilbert space dimension $n=C_{N/2}^{N}$). Throughout the paper, we set $J=1$. The sign of $J$ does not affect the calculations involving the entire spectrum, which is obtained from the full \textit{exact diagonalization} methods \cite{Debojyoti_paper_1_2022}. By varying the physical crossover parameters $h$ and $J_t$, the \textit{convensional} and \textit{unconvensional} time-reversal symmetries are systematically broken, we can achieve the Poissonian, GOE, and GUE statistics and crossovers amongst them (for details see the Ref. \cite{Debojyoti_paper_1_2022}).

\begin{figure}[h]
\begin{centering}
\includegraphics[scale=0.65]{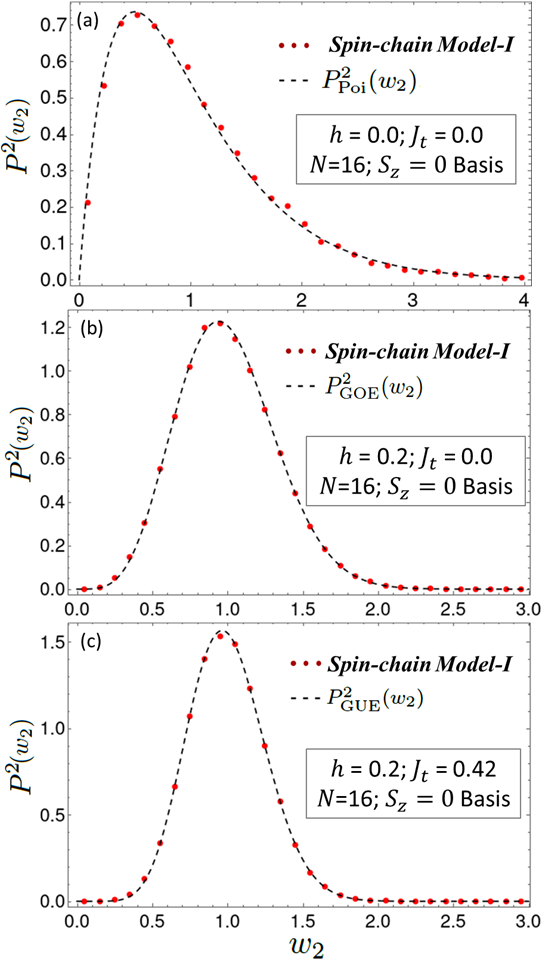}
\par\end{centering}
\caption{The nNNSD of the spin-chain model-I ($N=16$ system). (a) represents
the integrable limit ($h=0.0$ and $J_{t}=0.0$), which follows the
semi-Poissonian statistics {[}$P_{\mathrm{Poi}}^{2}(w_{2})$ in Table
\ref{tab:nNNSD_formulas_table}{]}. (b) and (c) represent the GOE
($h=0.2$ and $J_{t}=0.0$) and the GUE ($h=0.2$ and $J_{t}=0.42$)
limits, and follow the Wigner-surmise-like results, $P_{\mathrm{GOE}}^{2}(w_{2})$
and $P_{\mathrm{GUE}}^{2}(w_{2})$ (listed in Table \ref{tab:nNNSD_formulas_table}),
respectively. \label{fig:nNNSD_Poissonian-spin_chain}}
\end{figure}

\subsubsection{Model-II}

Model-II is represented by another half-filled, $N$ site, one-dimensional spin-$\frac{1}{2}$ Hamiltonian $H_2$, given by \cite{Debojyoti_paper_2_2023}
\begin{align} 
\nonumber H_2 & =H_{h}+H_{ir}+H_{r}+H_{DM} \\ \nonumber  & =\sum_{j=1}^{N-1}J\boldsymbol{\mathrm{S}}_{j}.\boldsymbol{\mathrm{S}}_{j+1}+\sum_{j=1}^{N-1}J\epsilon_{j}\mathrm{S^{z}}_{j}\mathrm{S^{z}}_{j+1}\\  & +\sum_{j=1}^{N}h_{j}\mathrm{S^{z}}_{j}+\sum_{j=1}^{N-1}\boldsymbol{\boldsymbol{D}}\cdot[\mathbf{S}_{j}\times\mathbf{S}_{j+1}].
\label{eq:model-II_Hamiltonian} 
\end{align} 
Among the four terms of this spin-chain Hamiltonian, $H_{h}$ and $H_{r}$ also appear in model-I. $H_{ir}$ is the second random term in model-II, which is called the random \textit{Ising} term, where the exchange interaction is randomized by multiplying $\ensuremath{J}$ with the dimensionless random parameter $\epsilon_{j}$, which follows a Gaussian distribution having zero mean and variance $\epsilon^{2}$. In a real system this could arise from a randomly varying lattice spacing due to quenched disorder, and a resultant random superexchange coupling between neighboring spins. $H_{DM}$ is the well known anti-symmetric \textit{Dzyaloshinskii-Moriya} (DM) interaction term \cite{Moriya_DM_1,Moriya_DM_2,Dzyaloshinskii_DM,Hamazaki_DM_RMT_paper,Vahedi_Quantum_chaos_DM}, describing the anisotropic effective spin-spin coupling between the neighboring spins. While competing with the Heisenberg term, the anti-symmetric spin part, ($\mathbf{S}_{j}\times\mathbf{S}_{j+1}$), results in \textit{canted spin arrangements} \cite{Yosida_Theory_of_Magnetism} in real materials.

Unlike $H_{1}$, $H_{2}$ does not commute with the total $\mathrm{S^{z}}$,
in general, due to the presence of $H_{DM}$, so we use the full $2^{N}$
number of basis states for some of our calculations involving model-II.
On the other hand, if only finite $H_{h}$ and $H_{ir}$ terms present
in a system, the total $\mathrm{S^{z}}$ remains conserved, and we
need to work in a constant to a $\mathrm{S^{z}}$ sector. Here we
choose an \textit{odd} value for $N$ to ensure that our spin-1/2
chain has a doubly degenerate eigenspectrum (in full basis calculations),
which is referred to as the \textit{Kramers degeneracy} (KD). By tuning
the physical crossover parameters $\epsilon,$ $h$, and $D$ (all
measured in units of $J$), we break several unitary (spin-rotational)
and anti-unitary (time-reversal) symmetries, and can achieve the Poissonian
and the three canonical Wigner-Dyson ensembles (for details see the
Ref. \cite{Debojyoti_paper_2_2023}). 

\subsubsection{Methodology and Results for the Spin-chain Models}

It is known that the density of states (DOS) of the RMT matrices follow
the classic Wigner's semi-circular form \cite{Mehta-Book}. However,
as is well known, the DOS of a quantum chaotic system follows a more
Gaussian-like distribution, due to a finite range of interactions
in a physical system resulting in much sparser matrices compared to
the RMT ensembles. However, surprisingly, the NNSD and other short-range
higher-order spectral correlations for physical systems seem to be
somewhat insensitive to this difference between physical systems and
RMT, and show excellent agreement between the RMT predictions (Wigner-surmises)
and the corresponding distributions from non-interacting as well as
interacting many-body systems \cite{Debojyoti_paper_1_2022,Debojyoti_paper_2_2023,Santhanam_review_QKR_2022,Kota_Embedded_Random_Matrix_Ensembles_billiards,Altshuler_book_quantum_transport_billiard,Abul_Magd_surmise_like_crossover,Avishai-Richert-PRB}.
As far as the nNNSD is concerned, we have already seen that the RMT
predicted results show very good agreement with those derived from
non-interacting quantum chaotic models like the QKR or the Chaotic
billiard systems. In this Section, we will discuss the comparison
of the RMT based predictions to the results from the above two interacting
quantum many-body models. 

When studying the correlation statistics of a physical system, it
is a common practice to only analyze the \textit{mid} of the spectrum
\cite{Nandkishore_Many-body_localization_and_thermalization_2016},
where the states are closest to being ergodic. In Refs. \cite{Debojyoti_paper_1_2022}
and \cite{Debojyoti_paper_2_2023}, we have already shown that for
a significantly large system size, we can consider the \textit{full}
spectrum and study the spectral correlation statistics of the disordered,
interacting quantum systems and this is what we consider here too.
Before studying the nNNSD, we perform standard unfolding \cite{Mehta-Book,Debojyoti_paper_1_2022,Debojyoti_paper_2_2023}
of the spectrum to eliminate any dependence on system-specific level-density.

To check the robustness of the Wigner-surmise-like nNNSD results obtained
from the RMT matrix models in Sec. \ref{sec:METHODOLOGY-and-CALCULATIONS},
we want to compare them against the corresponding nNNSD's of the eigenspectra
from both the interacting quantum spin models (model-I and model-II).
In the integrable limit ($h=0.0$ and $J_{t}=0.0$ \cite{Debojyoti_paper_1_2022})
of the $N=16$ system ($\mathrm{S^{z}}=0$ sector), following the
model-I Hamiltonian {[}Eq. (\ref{eq:model-I_Hamiltonian}){]}, we
compare its nNNSD with the RMT predicted semi-Poissonian statistics
{[}$P_{\mathrm{Poi}}^{2}(w_{2})${]} and plotted in Fig. \ref{fig:nNNSD_Poissonian-spin_chain}(a).
The nNNSD of model-I shows very good agreement with the RMT prediction.

\begin{figure}[H]
\begin{centering}
\includegraphics[scale=0.7]{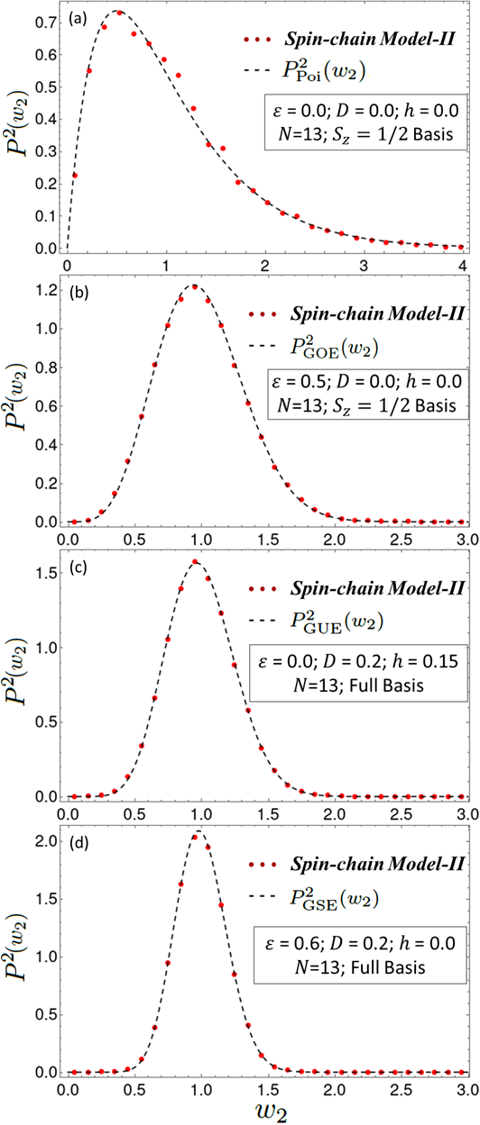}
\par\end{centering}
\caption{The nNNSD of the spin-chain model-II ($N=13$ system) at the GSE limit
($\varepsilon=0.6$ and $D=0.2$ \cite{Debojyoti_paper_2_2023}),
which follows the corresponding Wigner-surmise-like statistics {[}$P_{\mathrm{GSE}}^{2}(w_{2})$
in Table \ref{tab:nNNSD_formulas_table}{]}. \label{fig:nNNSD_GSE-spin_chain}}
\end{figure}

In Fig. \ref{fig:nNNSD_Poissonian-spin_chain}(b), we compare the
nNNSD of the model-I at the GOE regime ($h=0.2$ and $J_{t}=0.0$
for $N=16$ system in the $\mathrm{S^{z}}=0$ sector \cite{Debojyoti_paper_1_2022})
against the Wigner-surmise-like distribution $P_{\mathrm{GOE}}^{2}(w_{2})$.
Again, for the $N=16$ system ($\mathrm{S^{z}}=0$ sector), model-I
achieves the GUE distribution at $h=0.2$ and $J_{t}=0.42$ \cite{Debojyoti_paper_1_2022}.
The nNNSD of the eigenspectrum is compared with the corresponding
distribution $P_{\mathrm{GUE}}^{2}(w_{2})$ and plotted in Fig. \ref{fig:nNNSD_Poissonian-spin_chain}(c).

The NNSD of the model-II Hamiltonian {[}Eq. (\ref{eq:model-II_Hamiltonian}){]}
with $N=13$, follows the standard GSE distribution at $\varepsilon=0.6$,
$D=0.2$, and $h=0.0$ (full basis) \cite{Debojyoti_paper_2_2023}.
After removing the Kramers degeneracy from the eigenspectrum, we compare
the calculated nNNSD with the previously obtained Wigner-surmise-like
distribution $P_{\mathrm{GSE}}^{2}(w_{2})$, and plot them in Fig.
\ref{fig:nNNSD_GSE-spin_chain}(d). GUE is achieved in a scenario
when both \textit{conventional} and \textit{unconventional} time-reversal
symmetries are broken, \textit{i.e.}, at $\varepsilon=0.0$, $D=0.2$,
and $h=0.15$ \cite{Debojyoti_paper_2_2023}, the corresponding nNNSD
is plotted in Fig. \ref{fig:nNNSD_GSE-spin_chain}(c). For the completeness
of our studies involving the model-II, we also plot the nNNSD of the
semi-Poissonian {[}Fig. \ref{fig:nNNSD_GSE-spin_chain}(a){]} and
the GOE statistics {[}Fig. \ref{fig:nNNSD_GSE-spin_chain}(b){]},
achieved at $\varepsilon=0.0$ and $\varepsilon=0.5$ (fixed $D=0.0$
and $h=0.0$) \cite{Debojyoti_paper_2_2023}, respectively, while
working in the $\mathrm{S^{z}}=1/2$ sector. 

The excellent agreements between the nNNSD's of the single- and many-body
quantum chaotic systems (QKR, Sinai billiard, and spin-chain models
I and II; illustrated in Figs. \ref{fig:floquet_QKR_nNNSD}, \ref{fig:Sinai_Billiard_nNNSD},
\ref{fig:nNNSD_Poissonian-spin_chain}, and \ref{fig:nNNSD_GSE-spin_chain},
respectively) and the deduced nNNSD results (as presented in Table
\ref{tab:nNNSD_formulas_table}), conclusively demonstrate that the
standard Gaussian random matrix ensembles (GOE, GUE, and GSE) effectively
capture the short-range spectral correlation statistics of the quantum
chaotic systems.

\section{Interesting features in GOE-to-GUE and GSE-to-GUE crossovers\label{sec:Interesting-features-in-crossovers}}

In this section, we explore two distinct crossovers in the nNNSD domain:
one transitioning from the GOE to the GUE, and the other from the
\textit{non-standard} or \textit{diluted} GSE (as discussed in reference
\cite{Debojyoti_paper_2_2023}) to the GUE, using both crossover
matrix models and the above spin-chain models. Additionally, the striking
similarities between these two crossovers are discussed.

\subsubsection{GOE-to-GUE Crossover in nNNSD}

\begin{figure}[h]
\begin{centering}
\includegraphics[scale=0.35]{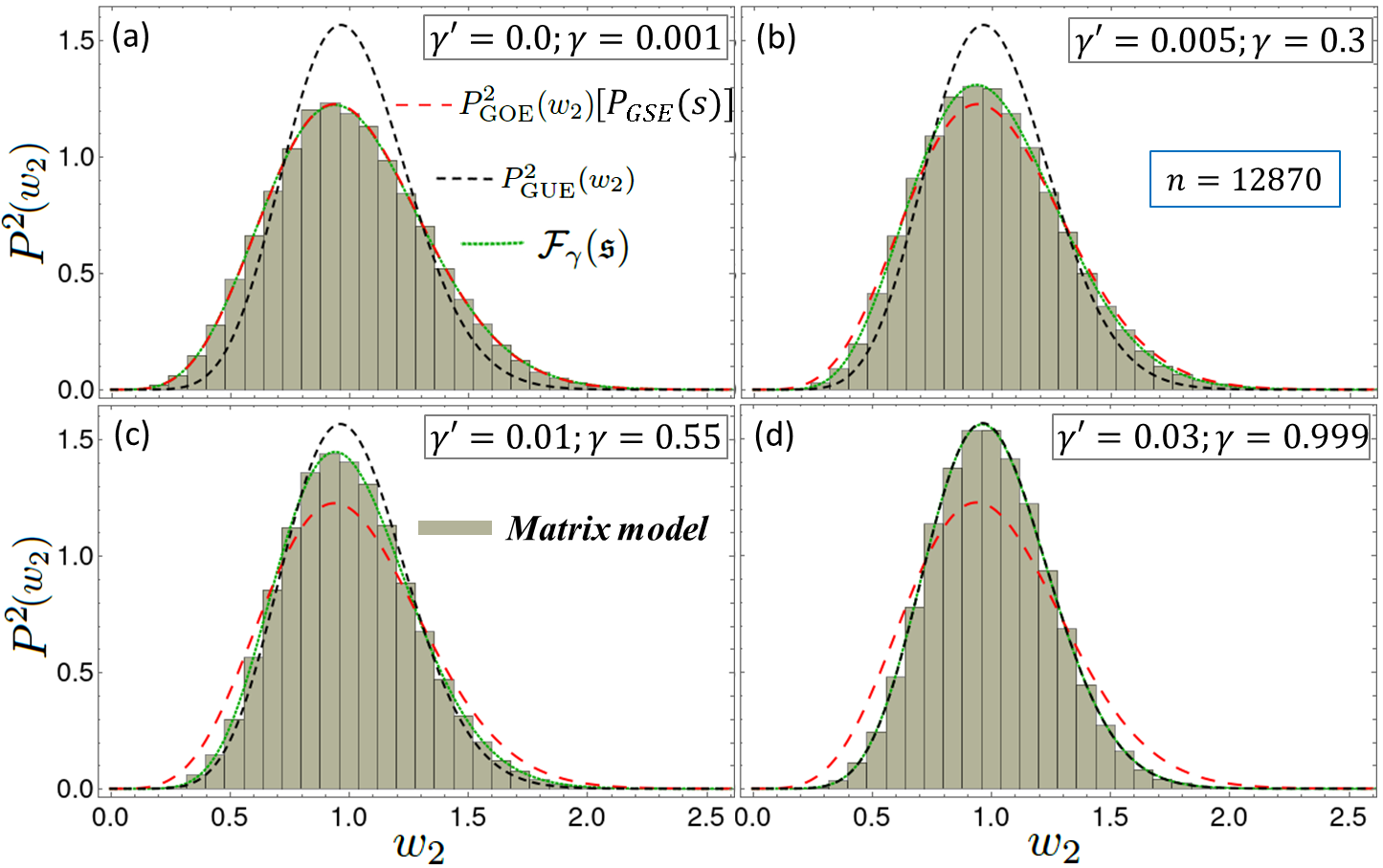}
\par\end{centering}
\caption{The nNNSD of the standard GOE-to-GUE crossover for the corresponding
matrix model $\mathcal{H}_{\mathrm{GOE-GUE}}$ {[}Eq. (\ref{eq:crossover_matrix_model_2}){]},
with a large dimension $n=12870$ (ensemble of $\mathcal{M}=5$ matrices).
We also observe that, the standard nNNSD of GOE, $P_{\mathrm{GOE}}^{2}(w_{2})$,
is equivalent to the standard NNSD of GSE, $P_{\mathrm{GSE}}(s)$
\cite{Mehta-Book}. In (a)-(d), we vary the effective crossover
parameter $\gamma'$ from $0.0$ to $0.03$. The crossover is also
compared against the derived semi-analytical interpolating distribution
$\mathcal{F}_{\gamma}(\mathfrak{s})$ with increasing crossover parameter
($\gamma$) related to the three-dimensional JPDF. \label{fig:GOE-GUE_nNNSD_crossover_matrix_model}}
\end{figure}

At first, we consider the crossover matrix model for the GOE-to-GUE
crossover, $\mathcal{H}_{\mathrm{GOE-GUE}}$ {[}Eq. (\ref{eq:crossover_matrix_model_2}){]}
with dimension $n=12870$ (chosen to mimic the dimension of the relevant
subspace in our spin model, see below) and an ensemble of $\mathcal{M}=5$
matrices, to study the crossover in nNNSD. In Fig. \ref{fig:GOE-GUE_nNNSD_crossover_matrix_model},
we plot the nNNSD for increasing \textit{effective} crossover parameter
$\gamma'$ of this large $n$ matrix model (different from the crossover
parameter $\gamma$ for a $3\times3$ matrix). Now it so turns out
that the nNNSD for the GOE class and the NNSD for the GSE class possess
equivalent statistical behavior \cite{Mehta-Book}, so we can effectively
observe the crossover between two distinct spectral correlation statistics
(nNNSD and NNSD) by tuning the crossover parameter $\gamma'$. In
Fig. \ref{fig:GOE-GUE_nNNSD_crossover_matrix_model}(a)-(d), we trace
this crossover by fitting with the semi-analytical interpolating distribution
$\mathcal{F}_{\gamma}(\mathfrak{s})$ (derived using the $3\times3$
dimensional crossover matrix model in Sec. \ref{subsec:Derivation-of-exact-analyrical-GOE-to-GUE_crossover})
with increasing values of the RMT crossover parameter $\gamma$, and
compare against the effective crossover parameter $\gamma'$. We observe
that at the GUE crossover, the parameter $\gamma\sim1$, whereas $\gamma'$
takes a much smaller value $0.03$. The extreme differences in the
values of the two crossover parameters ($\gamma$ and $\gamma'$)
are expected due to scaling effects arising from the differences in
their parent matrix dimensions, $3$ and $12870$, respectively, as
previously observed in Ref. \cite{Debojyoti_paper_2_2023}. 

\begin{figure}[h]
\begin{centering}
\includegraphics[scale=0.35]{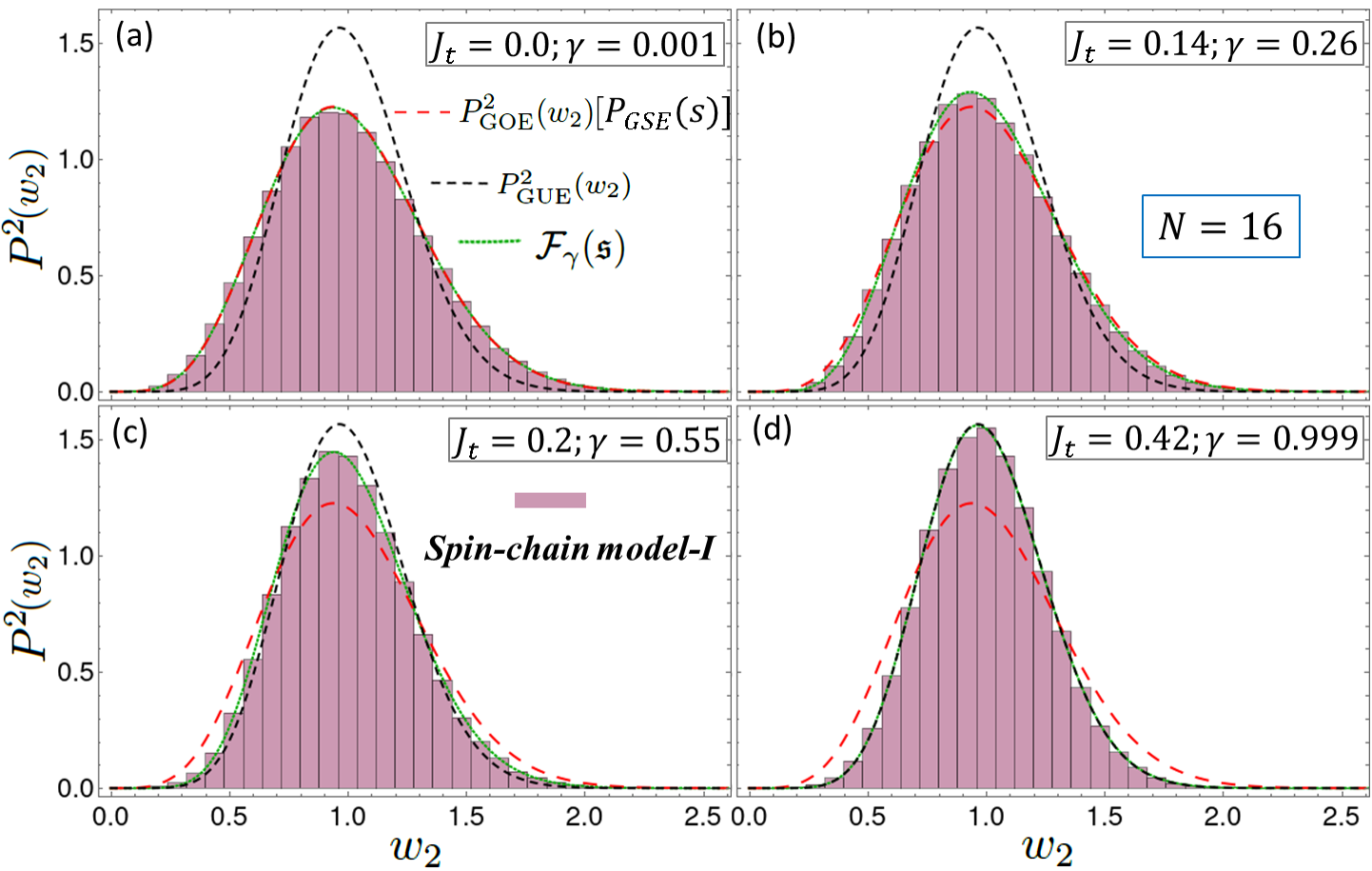}
\par\end{centering}
\caption{The nNNSD of the standard GOE-to-GUE crossover for the disordered
spin model-I with $N=16$ ($n=12870$). In (a)-(d), we vary the physical
crossover parameter $J_{t}$ from $0.0$ to $0.42$, by keeping $h$
fixed at $0.2$ \cite{Debojyoti_paper_1_2022}. The crossover is also
compared against the derived semi-analytical interpolating distribbution
$\mathcal{F}_{\gamma}(\mathfrak{s})$ with increasing crossover parameter
($\gamma$) related to the three-dimensional JPDF. \label{fig:GOE-GUE_nNNSD_crossover_spin_model_1}}
\end{figure}

In model-I, the GOE-to-GUE crossover is achieved by tuning the amplitude
of scalar chirality interaction, $J_{t}$, while keeping $h$ fixed
at the value of $0.2$ throughout, originally required to clamp the
system in the GOE regime. To establish the robustness of the above
crossover study, in Fig. \ref{fig:GOE-GUE_nNNSD_crossover_spin_model_1}(a)-(d),
we plot this nNNSD crossover in model-I with $N=16$ sites, and compare
the physical crossover parameter $J_{t}$ against the RMT crossover
parameter $\gamma$, corresponding to the semi-analytical interpolating
distribution $\mathcal{F}_{\gamma}(\mathfrak{s})$.

\subsubsection{GSE-to-GUE Crossover in nNNSD}

\begin{figure}[h]
\begin{centering}
\includegraphics[scale=0.35]{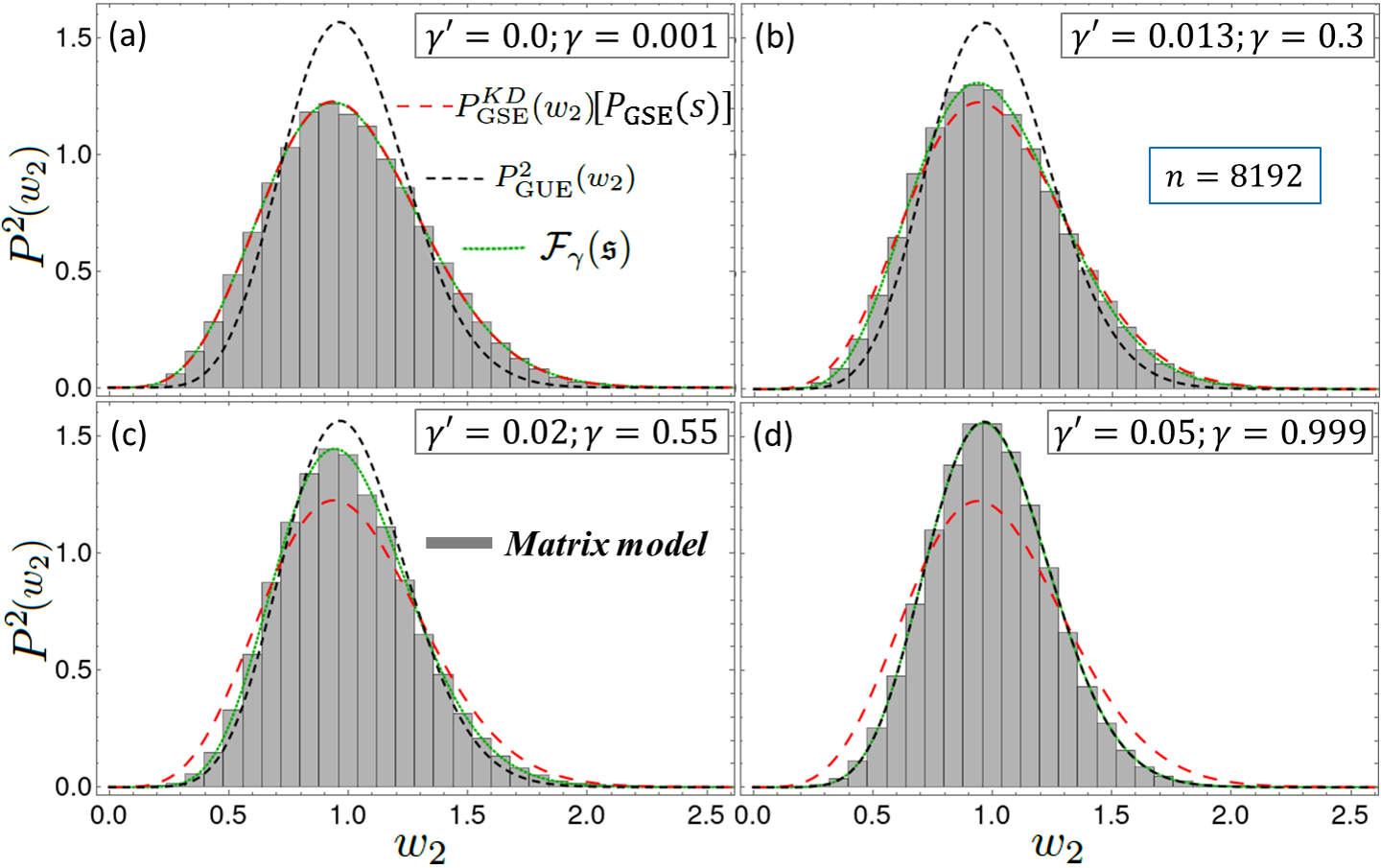}
\par\end{centering}
\caption{The nNNSD crossover between the \textit{diluted} GSE and the standard
GUE for the corresponding matrix model $\mathcal{H}_{\mathrm{GSE-GUE}}$
{[}Eq. (\ref{eq:crossover_matrix_model_2}){]}, with a large dimension
$n=8192$ (ensemble of $\mathcal{M}=10$ matrices). We observe that,
at the GSE limit, the\textit{ diluted} nNNSD, $P_{\mathrm{GSE}}^{KD}(w_{2})$,
is equivalent to the standard NNSD $P_{\mathrm{GSE}}(s)$. In (a)-(d),
we vary the effective crossover parameter $\gamma'$ from $0.0$ to
$0.05$. The crossover is also compared against the derived semi-analytical
interpolating distribution $\mathcal{F}_{\gamma}(\mathfrak{s})$ with
increasing crossover parameter ($\gamma$) related to the three-dimensional
JPDF. \label{fig:GSE-GUE_nNNSD_crossover_matrix_model}}
\end{figure}

While studying spectral correlation statistics of the standard GSE,
it is customary to remove one of the eigenvalues from each of the
Kramers doublets, whereas interesting statistics is observed in the
\textit{diluted} GSE scenario, where the Kramers degeneracy is retained
\cite{Debojyoti_paper_2_2023}. In this context, we investigate
the nNNSD crossover between the \textit{diluted} GSE and the standard
GUE symmetry classes. To study this crossover, we consider the corresponding
form of the crossover matrix model of Eq. (\ref{eq:crossover_matrix_model_1}),
\begin{equation}
\mathcal{H}_{\mathrm{GSE-GUE}}=\sqrt{1-\gamma'^{2}}\,\mathcal{H}_{\mathrm{GSE}}+\gamma'\,\mathcal{H}_{\mathrm{GUE}},\label{eq:GSE-GUE_crossover-matrix_model}
\end{equation}
 with $n=8192$ and an ensemble of $\mathcal{M}=10$ matrices, where
$\gamma'$ is the \textit{effective} crossover parameter for such
a large matrix dimension. We observe that, in the limit $\gamma'=0.0$,
the nNNSD\textit{ }of the \textit{diluted} GSE, denoted as $P_{\mathrm{GSE}}^{KD}(w_{2})$,
follows the NNSD of the \textit{standard} GSE, $P_{\mathrm{GSE}}(s)$
{[}Figs. \ref{fig:GSE-GUE_nNNSD_crossover_matrix_model}(a){]}. This
is to be expected, as we do not encounter the zero spacings from the
Kramers doublets while studying the next nearest-neighbor spacings,
leaving us with only the effective nearest-neighbor spacings in the
GSE regime \cite{Debojyoti_paper_2_2023}. Now, for a finite $\gamma'$,
the KD is lifted from the eigenspectrum, and by increasing $\gamma'$,
we study the GSE-to-GUE crossover in nNNSD {[}Figs. \ref{fig:GSE-GUE_nNNSD_crossover_matrix_model}(b)-(d){]}.
At $\gamma'=0.05$ {[}Fig. \ref{fig:GSE-GUE_nNNSD_crossover_matrix_model}(d){]},
we observe that the distribution follows the nNNSD of the GUE, $P_{\mathrm{GUE}}^{2}(w_{2})$.
Therefore, by varying the parameter $\gamma'$ in the GSE-to-GUE crossover,
we are able to demonstrate a transition from one RMT correlation study,
the NNSD, to another, the nNNSD. We also plot the semi-analytical
interpolating distribution $\mathcal{F}_{\gamma}(\mathfrak{s})$ (discussed
in Sec. \ref{subsec:Derivation-of-exact-analyrical-GOE-to-GUE_crossover}),
and compare the two crossover parameters $\gamma'$ and $\gamma$.
Similar to the GOE-to-GUE crossover scenario, in this case as well,
we witness a substantial disparity in the values of the two crossover
parameters, owing to analogous matrix dimension considerations.

\begin{figure}[h]
\begin{centering}
\includegraphics[scale=0.35]{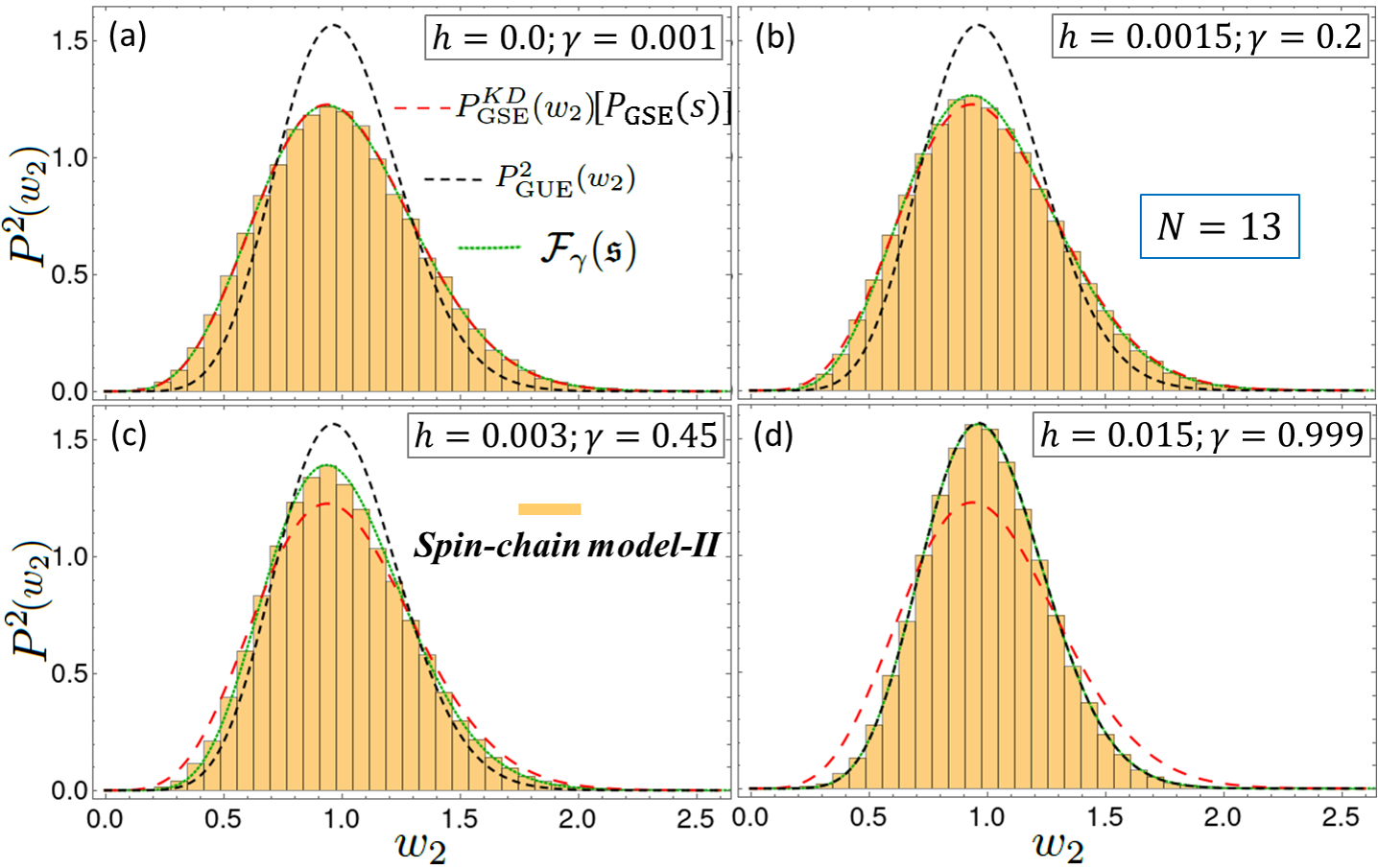}
\par\end{centering}
\caption{The nNNSD crossover between the \textit{diluted} GSE and the standard
GUE for the disordered spin model-II with $N=13$ ($n=8192$ and $\mathcal{M}=10$).
We observe that, at the GSE limit, the \textit{diluted} nNNSD, $P_{\mathrm{GSE}}^{KD}(w_{2})$,
is equivalent to the standard NNSD $P_{\mathrm{GSE}}(s)$. In (a)-(d),
we vary the physical crossover parameter $h$ from $0.0$ to $0.015$
by keeping fixed parameters at $\epsilon=0.6$ and $D=0.2$ \cite{Debojyoti_paper_2_2023}.
The crossover is also compared against the derived semi-analytical
interpolating distribution $\mathcal{F}_{\gamma}(\mathfrak{s})$ with
increasing crossover parameter ($\gamma$) related to the three-dimensional
JPDF. \label{fig:GSE-GUE_nNNSD_crossover_spin_model_2}}
\end{figure}

To establish the generality of this interesting crossover, we investigate
a similar crossover in our spin-chain model-II with lattice size $N=13$.
Now, without removing KD from the spectra, we plot the nNNSD of the
system in the \textit{diluted} GSE limit ($\epsilon=0.6$, $D=0.2,$
and $h=0.0$ from the Ref. \cite{Debojyoti_paper_2_2023}) in Fig.
\ref{fig:GSE-GUE_nNNSD_crossover_spin_model_2}(a). As expected, it
is identical to the standard NNSD $P_{\mathrm{GSE}}(s)$. For any
finite $h$, the KD is lifted, and we study the \textit{diluted} GSE-to-GUE
crossover in nNNSD, by tuning $h$ {[}Figs. \ref{fig:GSE-GUE_nNNSD_crossover_spin_model_2}(b)-(d){]}.
At $h=0.015$ {[}same value at which NNSD crossover is observed in
Ref. \cite{Debojyoti_paper_2_2023}{]}, we achieve the nNNSD of
GUE {[}Fig. \ref{fig:GSE-GUE_nNNSD_crossover_spin_model_2}(d){]}.
In Fig. \ref{fig:GSE-GUE_nNNSD_crossover_spin_model_2}, we observe
that the semi-analytical interpolating distribution $\mathcal{F}_{\gamma}(\mathfrak{s})$,
which is characterized by an increasing RMT crossover parameter $\gamma$,
effectively captures the limiting and intermediate cases of this physical
crossover. Also, a comparison is drawn between the values of the physical
symmetry-breaking parameter $h$ and the RMT crossover parameter $\gamma$.

Although both of these nNNSD crossovers (the transition from the standard
GOE to the standard GUE and from the diluted GSE to the standard GUE),
exhibit distinct symmetry settings corresponding to the two distinct
spin-chain models, it is noteworthy that, they can be conceptualized
in terms a unified mathematical expression for the distribution functions
that characterize the effective RMT transitions. This
understanding arises from the fact that, the behavior of nNNSD in
the standard GOE regime effectively mirrors the NNSD of the standard
GSE \cite{Mehta-Book}. Additionally, this equivalence holds true
when considering every \textit{second} 
eigenvalue for calculating the NNSD (or effectively the nNNSD) in
the case of the \textit{diluted} GSE. Thus, both crossovers start from a statistically equivalent point
and eventually propagate to the standard GUE regime. Based on the
analysis of the Figs. \ref{fig:GOE-GUE_nNNSD_crossover_matrix_model}
and \ref{fig:GSE-GUE_nNNSD_crossover_matrix_model}, it is evident
that the effective crossover parameters $\gamma'$ associated with
large-dimensional ($n=12870$ and $8192$, respectively) crossover
matrix models is considerably smaller than the corresponding crossover
parameters $\gamma$, linked to the semi-analytical interpolating
distribution $\mathcal{F}_{\gamma}(\mathfrak{s})$, based on $3\times3$
matrix models. The observed phenomenon aligns with the understanding
presented in the Refs. \cite{PandeyMehta1983,KumarPandey2011b,Debojyoti_paper_1_2022}
that, as the matrix dimensions increase (linked monotonically to the
size of a physical system, for physical Hamiltonians), a smaller crossover
parameter value can induce a transition from one RMT ensemble to another,
thus gradually abolishing the lag seen in the symmetry-breaking, as
a function of the mixing or symmetry-breaking parameter, for finite
systems.

\section{Conclusions\label{sec:Conclusions}}

In this paper, we have explored the {\em Wigner-surmise-like} and {\em exact
analytical} results for the next-nearest-neighbor spacing distributions associated with 
random matrix models, whose spectral correlation statistics belong
to the integrable class (semi-Poissonian) or various Wigner-Dyson symmetry classes (GOE, GUE, and GSE). 
The extended-Wigner-surmise results were deduced by stringent fittings to relevant spectral fluctuation statistics of large random matrices with appropriate symmetries.
On the other hand, the exact analytical results for the GOE and GUE classes were derived by considering the joint probability density of eigenvalues from the $3\times3$ matrix models.
%\cite{Porter_nNNSD_GOE_1963,Porter_nNNSD_GUE_1963,Porter_nNNSD_book_1965,PandeyMehta1983,MehtaPandey1983,Mehta-Book,Ayana's_PhysRevE},
%and implementing the relevant change of variable methods \cite{Porter_nNNSD_GOE_1963,Porter_nNNSD_GUE_1963,Porter_nNNSD_book_1965,AtasPRL2013_ratio_1,Atas_2013_ratio_2,Ayana's_PhysRevE},
%we deduce the exact analytical nNNSD results for the
%Gaussian orthogonal and unitary ensembles.
Moreover, using the analytical expression
of the JPDF for the GOE-to-GUE crossover, we have derived the corresponding
semi-analytical distribution 
%{[}$\mathcal{F}_{\gamma}(\mathfrak{s})${]} 
that interpolates between GOE and GUE
nNNSD's. Our study also involved examining the concurrence between
our nNNSD results and the spectral statistics of both single-body
(QKR and QQSB) and many-body (spin-chain model-I and model-II) quantum
chaotic systems. The key findings of this paper are listed below:

(1) Our findings reveal that the effective Dyson indices ($\beta'$)
associated with the nNNSD exhibit a robust behavior, in the sense that the $\beta'$
values depend solely on the Wigner-Dyson symmetry classes, rather than on the specific statistical
measures of correlation, such as nNNSD or {\em second order} RD. 

(2) The {\em Wigner-surmise-like} nNNSD results thus obtained, accurately capture the spectral correlation
statistics of the studied quantum chaotic systems, {\em viz.} the single-body
Quantum Kicked Rotor and Quantum Billiard systems, as well as the many-body disordered
spin-chain model-I and model-II, both in the {\em integrable} and the {\em quantum chaotic} limits.

(3) The {\em exact analytical} nNNSD results, obtained from the three-dimensional
joint probability density of eigenvalues for the GOE and the GUE, demonstrate
a harmonious correspondence with the semi-Poissonian or the Wigner-surmise-like nNNSD findings,
as well as with the corresponding nNNSD of the single-body and many-body
quantum chaotic systems, in the {\em integrable} and the {\em chaotic} limits, respectively. 
This, in a way, validates the BGS and the Berry-Tabor conjectures for the nNNSD, going beyond NNSD.

(4) The variations of the derived crossover semi-analytical nNNSD form {[}$\mathcal{F}_{\gamma}(\mathfrak{s})${]}, interpolating between the GOE and the GUE, as a function of the crossover parameter $\gamma$, show excellent agreement with the transition
between the two limiting cases (GOE and GUE), be it in the case of higher dimensional matrix models, or the single-body and many-body physical models, discussed in this paper.

(5) Finally, our analysis of two interesting nNNSD crossovers, the GOE-to-GUE transition and the {\em diluted}
GSE-to-GUE transition, in the context of both crossover matrix models and the above two interacting spin models (model-I and model-II), 
shows that both of these can be alternately viewed as crossovers between the NNSD of an alternate
statistics (the \textit{standard} GSE) and the nNNSD of the GUE, as far as their mathematical forms are concerned.

\begin{acknowledgments} D.K. and S.S.G. acknowledge the Science and Engineering Research Board (SERB), Department of Science and Technology (DST), Government of India, for financial support via Project No. ECR/2016/002054. S.K. acknowledges financial support provided by SERB, DST, Government of India, via Project No. CRG/2022/001751. D.K. acknowledges Ashutosh Dheer for his  assistance in gaining proficiency with the Kwant package. \end{acknowledgments}

\appendix

\begin{widetext}

\section{Derivation of \texorpdfstring{\MakeLowercase{n}NNSD}{nNNSD} for GOE and GUE from Three-dimensional
JPDF's}\label{Appendix}

In this appendix, we consider the three-dimensional joint probability
densities of the Gaussian Orthogonal and Gaussian Unitary matrix ensembles
and following the similar methods implemented in the Refs. \cite{Porter_nNNSD_book_1965,Porter_nNNSD_GOE_1963,Porter_nNNSD_GUE_1963},
we derive the exact nNNSD expressions presented Eqs. (\ref{eq:normalized_analytical_GOE})
and (\ref{eq:normalized_analytical_GUE}), which follow both the normalization
and unit mean next nearest-neighbor spacing conditions. The three-dimensional
joint probability densities for the GOE and GUE are given by $q_{\mathrm{GOE}}(x_{1},x_{2},x_{3})$
and $q_{\mathrm{GUE}}(x_{1},x_{2},x_{3})$, respectively, and represented
as \cite{PandeyMehta1983,MehtaPandey1983,Mehta-Book,Ayana's_PhysRevE},
\begin{equation}
q_{\mathrm{GOE}}(x_{1},x_{2},x_{3})=\frac{1}{48\sqrt{2}\pi v^{6}}e^{-\frac{1}{4v^{2}}(x_{1}^{2}+x_{2}^{2}+x_{3}^{2})}\left|(x_{1}-x_{2})(x_{1}-x_{3})(x_{2}-x_{3})\right|,\label{eq:GOE_analytical_JPDF}
\end{equation}
\begin{equation}
q_{\mathrm{GUE}}(x_{1},x_{2},x_{3})=\frac{1}{768\pi^{3/2}v^{9}}e^{-\frac{1}{4v^{2}}(x_{1}^{2}+x_{2}^{2}+x_{3}^{2})}\left[(x_{1}-x_{2})(x_{1}-x_{3})(x_{2}-x_{3})\right]^{2},\label{eq:GUE_analytical_JPDF}
\end{equation}
 where $v$ is the corresponding scaling parameter. For convenience,
we consider $v=1$ throughout our calculations. Arranging the eigenvalues
in the ascending order, $x_{1}\leqslant x_{2}\leqslant x_{3}$, we
consider the next nearest-neighbor spacing as $\mathfrak{s}=x_{3}-x_{1}$,
we will impose the unit mean level spacing condition later and accordingly
scale $\mathfrak{s}$. The corresponding probability density function
for the GOE is obtained as 
\begin{equation}
Q_{\mathrm{GOE}}(\mathfrak{s})=3!\int_{-\infty}^{\infty}dx_{2}\int_{-\infty}^{x_{2}}dx_{1}\int_{x_{2}}^{\infty}dx_{3}\;q_{\mathrm{GOE}}(x_{1},x_{2},x_{3})\delta\left[\mathfrak{s}-(x_{3}-x_{1})\right].\label{eq:analytical_general_ensemble_probability_density_1}
\end{equation}
 Using the similar methods used in Refs. \cite{AtasPRL2013_ratio_1,Atas_2013_ratio_2,Ayana's_PhysRevE}
for the ratio distribution, we implement the change of variables ($x_{1},x_{2},x_{3}$)
$\rightarrow$ ($x,x_{2},y$), by considering $x=x_{2}-x_{1}$ and
$y=x_{3}-x_{2}$, and get 
\begin{equation}
Q_{\mathrm{GOE}}(\mathfrak{s})=6\int_{-\infty}^{\infty}dx_{2}\int_{-\infty}^{x_{2}}dx\int_{x_{2}}^{\infty}dy\;q_{\mathrm{GOE}}(x_{2}-x,x_{2},x_{2}+y)\delta\left[\mathfrak{s}-(x+y)\right].\label{eq:analytical_general_ensemble_probability_density}
\end{equation}
 Using Eq. (\ref{eq:GOE_analytical_JPDF}), we first trivially perform
the integral over $x_{2}$ and left with 
\begin{equation}
Q_{\mathrm{GOE}}(\mathfrak{s})=\frac{1}{4\sqrt{6\pi}}\int_{0}^{\infty}dx\int_{0}^{\infty}dy\;e^{-\frac{(x^{2}+xy+y^{2})}{6}}xy(x+y)\delta\left[\mathfrak{s}-(x+y)\right].\label{eq:analytical_GOE_1}
\end{equation}
 Now, performing the integrals over $y$ and $x$, we get the probability
density 
\begin{equation}
Q_{\mathrm{GOE}}(\mathfrak{s})=\frac{1}{16\sqrt{6\pi}}\exp\left(-\mathfrak{s}^{2}/6\right)\mathfrak{s}\left(12\mathfrak{s}+\exp\left(\mathfrak{s}^{2}/24\right)\sqrt{6\pi}\left(\mathfrak{s}^{2}-12\right)\mathrm{erf}\left[\frac{\mathfrak{s}}{2\sqrt{6}}\right]\right),\label{eq:analytical_GOE_2}
\end{equation}
 where $\mathrm{erf}\left[z\right]=\frac{2}{\sqrt{\pi}}\int_{0}^{z}e^{-t^{2}}dt$
is the \textit{error function}. We check that, $Q_{\mathrm{GOE}}(\mathfrak{s})$,
follows the normalization condition $\int_{0}^{\infty}Q_{\mathrm{GOE}}(\mathfrak{s})d\mathfrak{s}=1$,
but the mean spacing is, $\int_{0}^{\infty}\mathfrak{s}Q_{\mathrm{GOE}}(\mathfrak{s})d\mathfrak{s}=3\sqrt{\frac{6}{\pi}}$.
After scaling $\mathfrak{s}$, so that the mean spacing is one, we
obtain the normalized probability density for GOE as
\begin{equation}
\mathcal{Q}_{\mathrm{GOE}}(\mathfrak{s})=3\sqrt{\frac{6}{\pi}}Q_{\mathrm{GOE}}\left(3\sqrt{\frac{6}{\pi}}\mathfrak{s}\right)=\frac{3^{4}\exp\left(-9\mathfrak{s}^{2}/\pi\right)\mathfrak{s}\left(6\mathfrak{s}-\exp\left(9\mathfrak{s}^{2}/4\pi\right)\left(2\pi-9\mathfrak{s}^{2}\right)\mathrm{erf}\left[\frac{3\mathfrak{s}}{2\sqrt{\pi}}\right]\right)}{4\pi^{2}}.\label{eq:normalized_analytical_GOE-Appendix}
\end{equation}

Now, using $q_{\mathrm{GUE}}(x_{1,}x_{2},x_{3})$ {[}Eq. (\ref{eq:GUE_analytical_JPDF}){]}
in the Eq. (\ref{eq:analytical_general_ensemble_probability_density})
and performing similar processes, we get the normalized probability
density for GUE, 
\begin{equation}
\mathcal{Q}_{\mathrm{GUE}}(\mathfrak{s})=\frac{3^{11}\sqrt{3}\exp\left(-3^{5}\mathfrak{s}^{2}/2^{4}\pi\right)\mathfrak{s}^{2}\left(2^{3}3^{3}\mathfrak{s}\left(-2^{5}\pi+3^{4}\mathfrak{s}^{2}\right)+\sqrt{3}\exp\left(3^{5}\mathfrak{s}^{2}/2^{6}\pi\right)\left(2^{10}\pi^{2}-2^{6}3^{4}\pi\mathfrak{s}^{2}+3^{9}\mathfrak{s}^{4}\right)\mathrm{erf}\left[\frac{9}{8}\sqrt{\frac{3}{\pi}}\mathfrak{s}\right]\right)}{2^{20}\pi^{4}},\label{eq:normalized_analytical_GUE-Appendix}
\end{equation}
 which also follows the unit mean nearest-neighbor spacing condition.
It is to note that, after scaling, the spacing parameter $\mathfrak{s}$,
formally becomes equivalent to the parameter $w_{2}$.
\end{widetext}

\end{document}